\documentclass[referee]{raa}
\usepackage{graphicx,times}
\usepackage{natbib}
\usepackage{amssymb,amsmath}
\bibpunct{(}{)}{;}{a}{}{,}

\usepackage[a4paper=true,dvipdfm=true,pagebackref=true]{hyperref}
\hypersetup{pdftitle = The title of my PDF, pdfauthor = My name, pdfsubject= The subject, pdfkeywords = keyword1 keyword2 keyword3}
\hypersetup{colorlinks = true, linkcolor = green, anchorcolor = red, citecolor = blue, filecolor = red, pagecolor = red, urlcolor = red}

\usepackage{acronym, multirow}

\newcommand{\lsim}{\mathrel{\hbox{\rlap{\lower.55ex\hbox{$\sim$}} \kern-.3em \raise.4ex \hbox{$<$}}}}
\newcommand{\gsim}{\mathrel{\hbox{\rlap{\lower.55ex\hbox{$\sim$}} \kern-.3em \raise.4ex \hbox{$>$}}}}
\newcommand{\beq}{\begin{equation}}
\newcommand{\eeq}{\end{equation}}

\newcommand{\bes}{\begin{equation*}}
\newcommand{\ees}{\end{equation*}}
\newcommand{\beqa}{\begin{eqnarray}}
\newcommand{\eeqa}{\end{eqnarray}}
\newcommand{\bea}{\begin{eqnarray}}
\newcommand{\ena}{\end{eqnarray}}

\def\apjl{Astrophys. J. Lett.}
\def\apjs{Astrophys. J. Suppl.}

\def\aap{Astron. Astrophys.}

\acrodef{GW}{gravitational-wave}
\acrodef{EM}{electromagnetic}
\acrodef{BH}{black hole}
\acrodef{BBH}{binary black hole}
\acrodef{SBBH}{stellar-mass binary black hole}
\acrodef{BNS}{binary neutron star}
\acrodef{EMRI}{extreme mass-ratio inspiral}
\acrodef{MBBH}{massive black hole binary}
\acrodef{SNR}{signal-to-noise ratio}
\acrodef{AGN}{active galactic nucleus}
\acrodefplural{AGN}{active galactic nuclei}
\acrodef{SN Ia}[SN Ia]{type Ia supernova}
\acrodefplural{SN Ia}[SNe Ia]{type Ia supernovae}
\acrodef{PTA}{pulsar timing array}

\begin{document}

   \title{Gravitational-wave standard sirens and application in cosmology
}

 \volnopage{ {\bf 2012} Vol.\ {\bf X} No. {\bf XX}, 000--000}
   \setcounter{page}{1}

   \author{Wen Zhao\inst{1,2}, Liang-Gui Zhu\inst{1,2}, Youjun Lu\inst{3,4}
   }

   \institute{Department of Astronomy, University of Science and Technology of China, Hefei 230026, China; {\it wzhao7@ustc.edu.cn,~lianggui.zhu@ustc.edu.cn}\\
\and
School of Astronomy and Space sciences, University of Science and Technology of China, Hefei 230026, China\\ 
\and
School of Astronomy and Space Science, University of Chinese Academy of Sciences, Beijing 100049, China \\
\and
National Astronomical Observatories, Chinese Academy of Sciences, Beijing 100101, China {\it luyj@nao.cas.cn}\\ 
\vs \no
   {\small Received ??? ???? ??; accepted ??? ???? ??}
}

\abstract{
The discovery of the gravitational-wave event GW170817 from a binary neutron star merger, together with its multi-wavelength electromagnetic counterparts, marks the beginning of the era of multi-messenger gravitational-wave astronomy. Observations of gravitational-wave signals from compact binary mergers enable an independent measurement of the luminosity distance to the source. This implies that gravitational-wave sources can serve as “standard sirens” to probe the expansion history of the Universe, providing a new approach to constrain cosmological parameters. In this paper, we review the basic principles of using gravitational-wave standard sirens to constrain cosmology. We discuss various methods for determining the source distance and redshift, as well as the capabilities of second- and third-generation ground-based detectors and space-based detectors in constraining cosmological parameters, especially the Hubble constant and dark energy parameters. {By examining three types of standard sirens—binary neutron star mergers with electromagnetic counterparts as bright sirens,  stellar-mass binary black hole mergers as dark sirens, and the dark lensed sirens—we illustrate the methodology, challenges, and future prospects of the standard siren approach.}
\keywords{Cosmology: Hubble constant, dark energy --- gravitational-waves: standard sirens}
}

   \authorrunning{W. Zhao et al. }            
   \titlerunning{Gravitational-wave standard sirens and application in cosmology}  
   \maketitle

%
\section{Introduction}           
\label{sec1}

Einstein’s General Relativity is currently the most successful theory of gravity. In the more than one hundred years since its formulation, it has passed numerous experimental tests, including both laboratory experiments and astrophysical observations \citep{Test1,Test2,Test3,Test4}. Modern cosmology and high-energy astrophysics, which are based on General Relativity, have achieved tremendous success. In particular, the $\Lambda$CDM cosmological model can successfully explain almost all current cosmological observations and is therefore commonly referred to as the ``standard cosmological model'' \citep{Planck}.

Despite its great success, several fundamental problems remain unresolved in cosmology \citep{2025RSPTA.38340022E}. For example, the nature of dark energy and dark matter is still unknown. In particular, recent observations indicate several challenges for the standard $\Lambda$CDM model. One of the famous ones is the so-called Hubble tension: Different measurements yield significantly different values of the Hubble constant. The high-redshift measurements, including Cosmic Microwave Background radiation (CMB) and Baryon Acoustic Oscillations (BAO), always follow a smaller value of $H_0\sim 68\ {\rm km} \cdot{\rm s}^{-1}\cdot{\rm Mpc}^{-1}$ \citep{2020A&A...641A...6P, 2025JCAP...11..062L}, while the local distance-ladder determinations using Cepheid-calibrated supernovae always derive a larger value of $H_0\sim 73\ {\rm km} \cdot{\rm s}^{-1}\cdot{\rm Mpc}^{-1}$ \citep{2022ApJ...934L...7R}. The tension between the two measurements is approximately $5\sigma$ \citep{hubbletension}. Another important progress is on the dark energy: recent BAO observations from DESI project provide evidence for the evolution of the dark energy component at about $3\sigma$ level \citep{desi}. These issues indicate that our current understanding of the Universe may still be incomplete, and new 
probes are required to further test cosmological models and the theory of gravity itself.

The direct detections of \acp{GW} by the LIGO and Virgo collaborations have opened a new 
window to the Universe \citep{2016PhRvL.116f1102A}. Gravitational waves, as predicted by general relativity, are ripples in spacetime generated by accelerating massive objects, especially compact binary systems such as \acp{BNS} and \acp{BBH}. Unlike electromagnetic radiation, \acp{GW} propagate almost unimpeded through matter, carrying clean information about their sources.

One of the most remarkable features of \ac{GW} observations is that the waveform directly encodes the luminosity distance to the source \citep{1986Natur.323..310S}. This implies that \ac{GW} sources can serve as ideal “standard sirens”, analogous to standard candles in electromagnetic astronomy, but without relying on a cosmic distance ladder. By combining the luminosity distance measured from \acp{GW} with redshift information obtained through various methods, one can probe the expansion history of the Universe and constrain cosmological parameters \citep{2026SCPMA..6920401J}.

The detection of the \ac{BNS} merger event GW170817, together with its electromagnetic counterparts across multiple wavelengths, represents a milestone in this field \citep{2017Natur.551...85A}. It demonstrated, for the first time, the feasibility of using \ac{GW} standard sirens for cosmological measurements. Since then, \ac{GW} cosmology has developed rapidly and is expected to play an increasingly important role in the coming decades.

{In this paper, we review the basic principles and methodologies of using \ac{GW} standard sirens for cosmology. The key role of the method is to determine the luminosity distance and redshift of GW events independently. We review two methods proposed in the literature for measuring distances, i.e. by using the GW waveform and using strong gravitational lensing effects. For the measurement of redshift, we introduce seven methods proposed in the literature, including the methods using electromagnetic counterparts, using the redshift distribution of host galaxies or galaxy clusters, using the redshift distribution function of compact object binaries, using GW phase corrections due to tidal effects, using the mass distribution function of neutron stars, using the mass spectrum of stellar-mass black holes, using GW phase corrections from cosmic evolution. We discuss the advantages and disadvantages of different approaches, as well as the expected constraints on cosmological parameters from current and future \ac{GW} detectors. We also analyze three representative types of standard sirens and discuss their respective advantages, challenges, and prospects.}

\section{Gravitational-Wave Sources and Detectors}
\label{sec2}

There exists a wide variety of \ac{GW} sources in the Universe, and the frequencies and amplitudes of \acp{GW} generated by different sources vary significantly. Therefore, similar to the detection of electromagnetic waves, different methods have been developed to conduct full-band detection of \ac{GW} signals across various frequency bands. 

Currently, for high-frequency \ac{GW} sources, the mainstream international approach is to use laser interferometer \ac{GW} detectors. In particular, \ac{GW} sources in the $(1-10^{4})$ Hz band are primarily detected by ground-based laser interferometers. First-generation detectors, such as the Laser Interferometer Gravitational-wave Observatory (LIGO) \citep{1992Sci...256..325A}, Virgo \citep{2004CQGra..21S.385A}, GEO 600 \citep{2002CQGra..19.1377W}, and TAMA \citep{2004CQGra..21S.403T}, have already completed their missions. Second-generation detectors are currently operational, including the two Advanced LIGO detectors in the United States \citep{2015CQGra..32g4001L}, Advanced Virgo in Europe \citep{2015CQGra..32b4001A}, KAGRA in Japan \citep{2019NatAs...3...35K}, as well as LIGO-India \citep{2013IJMPD..2241010U}, which is under construction and expected to come online in the coming years. 
It was through the observations made by Advanced LIGO and Advanced Virgo that \ac{GW} emission from \ac{SBBH}, \ac{BNS} 
and neutron star-black hole (NSBH) mergers was first detected \citep{2016PhRvL.116f1102A, 2017PhRvL.119p1101A, 2021ApJ...915L...5A}, inaugurating the era of \ac{GW} astronomy. 

Meanwhile, third-generation ground-based \ac{GW} detectors are being designed worldwide and are anticipated to become operational around 2030. The two main proposals are the Einstein Telescope \citep[ET;][]{2010CQGra..27s4002P} in Europe and the Cosmic Explorer \citep[CE;][]{2017CQGra..34d4001A} in the United States. 
The former consists of six Michelson interferometers with arm lengths of 10 kilometers, arranged to form two equilateral triangles, effectively comprising three detectors. Its noise power spectral density is projected to be reduced by more than two orders of magnitude compared to second-generation detectors, and its low-frequency cutoff is expected to extend to approximately 1 Hz, significantly enhancing the detectable timescales of \acp{GW} and improving the detectors' localization capabilities. 
The latter will retain the ``L''-shaped detector configuration but extend the arm length to approximately 40 kilometers, building upon the second-generation detectors. While its noise performance is expected to surpass that of ET around $100$\,Hz, its low-frequency cutoff is only about $5$\,Hz. 
According to calculations, ET is projected to detect \ac{BNS} merger events up to a redshift of $z\sim 2$, whereas CE will achieve even greater detection depths \citep{2026JCAP...03..081A, 2021arXiv210909882E}. For \ac{SBBH} mergers, both detectors are expected to reach detection depths of up to $z \gtrsim 10$ \citep{2020JCAP...03..050M, 2021arXiv210909882E}. Therefore, third-generation ground-based \ac{GW} detectors are anticipated to usher in a new era of \ac{GW} cosmology, greatly enriching our understanding of the high-redshift universe.

\begin{figure}[!htb]
\begin{center}
\includegraphics[scale=0.8]{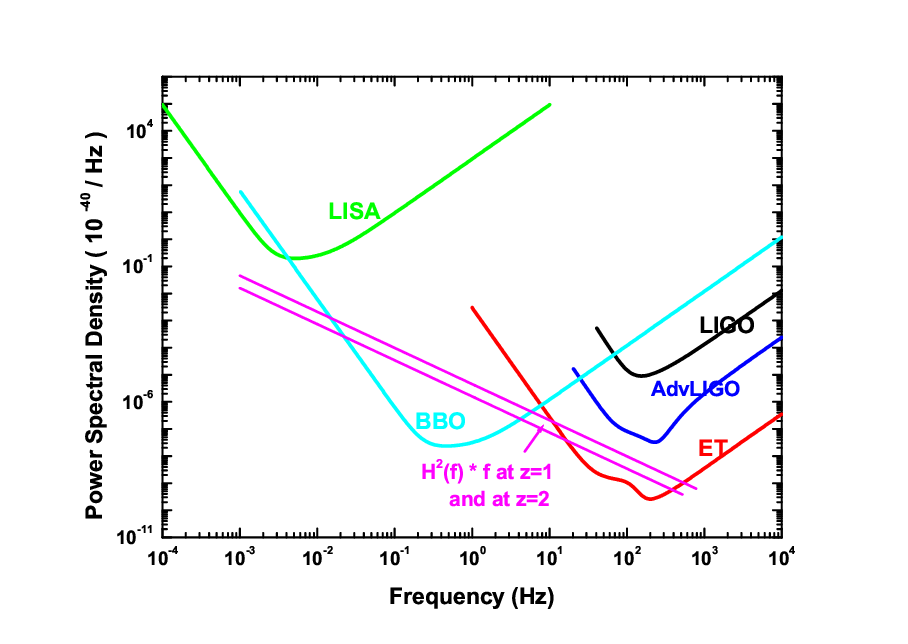}
\caption{Noise power spectral density curves of ground-based GW detectors (LIGO, Advanced LIGO, ET) and space-based GW detectors (LISA, BBO). The two pink solid lines represent the amplitude curves of the GW radiation from \ac{BNS} mergers at redshifts $z=1$ and $z=2$.}
\label{a1}
\end{center}
\end{figure}

For GW sources with somewhat lower frequencies $(10^{-4}-10^{0})$ Hz, space-based GW detectors have been conceived. The most mature project is the Laser Interferometer Space Antenna (LISA) mission, slated for launch around 2035 \citep{2017arXiv170200786A, 2024arXiv240207571C}. LISA will deploy three satellites in a heliocentric orbit to form a triangular constellation; laser links between the spacecraft will create a space-borne laser interferometer \ac{GW} observatory. With arm lengths on the order of millions of kilometers, LISA will be sensitive to \ac{GW} signals at much lower frequencies than ground-based instruments. Concurrently, China's TianQin \citep{2016CQGra..33c5010L, 2025CQGra..42q3001L} and Taiji \citep{2017NSRev...4..685H, 2022ChJSS..42..536L} projects are progressing steadily and are also expected to launch around 2035. 
The observational targets of space-based GW observatories are entirely different from those of ground-based detectors, primarily including \ac{MBBH} mergers, extreme mass-ratio inspirals (EMRIs), \ac{SBBH} inspirals, gravitational radiation from cosmic strings, and gravitational radiation from early universe phase transitions \citep{2020IJMPA..3550075R, 2023LRR....26....2A, 2023LRR....26....5A, 2025RPPh...88e6901L, 2023LRR....26....5A,2026LRR....29....1L}. Therefore, it is anticipated that space-based GW observatories will open up a new field in GW astronomy. 
In this paper, we mainly focus on LISA's observations of \acp{MBBH} and their cosmological applications. At the same time, more long-term space-based GW detector proposals have been put forward, mainly including the Big Bang Observer (BBO) \citep{2006CQGra..23.4887H} and the Deci-hertz Interferometer Gravitational-wave Observatory (DECIGO) \citep{2006CQGra..23S.125K}, with sensitive frequency bands mainly in $(10^{-3}-10^{4})$ Hz, particularly concentrated around $(10^{-1}-10^{1})$ Hz, thus complementing LISA and LIGO. 
In Figure \ref{a1}, we plot the noise curves and sensitive frequency bands of the first-generation LIGO, second-generation Advanced LIGO, and third-generation ground-based GW detector ET, as well as the space-based GW detectors LISA and BBO. For comparison, we also show the GW amplitudes of \ac{BNS} mergers, which are the main observational targets for ground-based interferometers and BBO. 
Note that, unlike electromagnetic wave observations, GW observations primarily measure the amplitude rather than the energy flux of GWs. Therefore, a reduction in instrument noise by one order of magnitude will increase the number of detectable sources by three orders of magnitude.

For extremely low-frequency GW signals $(10^{-9}-10^{-7})$ Hz, the primary detection method currently employed is the \ac{PTA} \citep{1990ApJ...361..300F}. This involves using radio telescopes or telescope arrays to monitor a set of millisecond pulsar signals over long periods, obtaining the distribution of timing residuals of these pulsars over time. By correlating the timing residuals of pulsars located in different spatial directions, one can search for the extremely low-frequency GW signals contained within them. Currently operational projects mainly include the Chinese CPTA (currently mainly FAST) \citep{2023RAA....23g5024X}, the American NANOGrav project \citep{2023ApJ...951L...8A}, the Australian PPTA project \citep{2023ApJ...951L...6R}, the European EPTA + Indian PTA project \citep[][]{2023A&A...678A..48E}, the MeerKAT PTA project (MPTA) \citep{2025MNRAS.536.1489M}, and the IPTA project \citep{2019MNRAS.490.4666P}, which combines NANOGrav, PPTA, and EPTA. After more than a decade of accumulated observations, these projects have all identified suspected signatures of a low-frequency background GW signal in their data \citep{2021ApJ...917L..19G,2021MNRAS.508.4970C,2020ApJ...905L..34A}. 
In the future, with further accumulation of observation time, and particularly with the inclusion of highly sensitive radio telescopes such as SKA \citep{2009IEEEP..97.1482D, 2015aska.confE..55M, 2015aska.confE.174B}, it is expected that the first detection of GW signals in this frequency band may be achieved in the near future, thereby opening a new observational window for GW detection. The primary GW sources in this frequency band are MBBHs in inspiral, including both individual sources and the GW background formed by a large number of sources. In addition, there are GW signals generated by cosmic strings during their motion, mergers, and collisions, which may originate from cosmic phase transitions, as well as primordial GW signals produced in the very early universe \citep{2019A&ARv..27....5B}. 

GW sources that can serve as ``standard sirens'' are currently primarily the merger or inspiral events of compact binary systems. These include \ac{BNS} mergers \citep{2017Natur.551...85A}, NSBH binary mergers \citep{2018PhRvL.121b1303V}, 
\ac{SBBH} mergers \citep{dark2, dark1}, MBBH inspirals or mergers \citep{2005ApJ...629...15H, 2011ApJ...732...82P, 2016JCAP...04..002T}, 
and EMRIs \citep{2008PhRvD..77d3512M, 2021MNRAS.508.4512L, 2024ApJS..273...24Z}. 
Among these, ground-based GW detectors mainly observe mergers of \ac{BNS}s, \acp{SBBH}, and NSBH binaries. Space-based detectors like LISA primarily observe \ac{MBBH} mergers and EMRIs \citep{2023LRR....26....5A}, while \acp{PTA} mainly observe the inspirals of MBBHs \citep{2020ApJ...889...79Y,PTA2}. For the gravitational radiation from such compact binary mergers, various theoretical models have been developed to accurately describe them \citep{2017ProgAstron..Chinese}. When the binary separation is large, the orbital velocities are non-relativistic, and the orbital decay due to GW radiation is slow, the post-Newtonian approximation can well describe the gravitational radiation. This stage is known as the inspiral phase. However, during the late inspiral and the merger phase, collectively referred to as the merger stage, the gravitational field becomes extremely strong, causing the post-Newtonian approximation to break down. Therefore, numerical relativity methods are typically employed to solve the equations. After the binary finally merges into a black hole, excess degrees of freedom are radiated away via GWs until a stationary black hole is formed. This stage is usually called the ringdown phase, and the emitted GWs can be described analytically using the quasinormal modes of black hole oscillations. Consequently, the GW waveform template for a binary merger event is effectively a superposition of these three parts, which is crucial for GW signal searches.

Based on joint observations by LIGO, Virgo, and KAGRA, the inferred merger rate for \acp{SBBH} is $(14-26) \, {\rm Gpc^{-3} \cdot yr^{-1}}$ \citep{ligo2025b}, which follows a yearly event rate of $(4.3-8.1) \, {\rm yr^{-1}}$ within the redshift range $z<0.1$, and $(8.1 \times 10^{3} - 1.5 \times 10^{4}) \, {\rm yr^{-1}}$ within $z<2$. The inferred merger rate for \ac{BNS}s is $(7.6 - 250) \, {\rm Gpc^{-3} \cdot yr^{-1}}$ \citep{ligo2025b}, which follows a yearly event rate of $(2.4 - 79) \, {\rm yr^{-1}}$ for $z<0.1$, and $(4.1 \times 10^{3} - 1.4 \times 10^{5}) \, {\rm yr^{-1}}$ for $z<2$. As the number of observed GW events gradually increases, estimates of their event rates will become increasingly precise. However, these estimates are not expected to change by orders of magnitude compared to the above figures. Therefore, we find that with third-generation GW detectors, the yearly number of compact binary merger events is projected to reach hundreds of thousands, or even millions, making large-sample statistical analysis feasible.


Both from theoretical models of galaxy formation and current observations of black holes, the cosmological-scale evolution and mergers of \acp{MBBH} are inevitable. These constitute the primary observational targets for future space-based GW detectors (such as LISA) and \acp{PTA}. However, current estimates of the merger rates for such events still carry significant uncertainties, involving many complex physical processes. \cite{2016PhRvD..93b4003K} conducted a detailed semi-analytical study of the event rates for \ac{MBBH} mergers under different galaxy evolution scenarios. For the formation of massive black holes, three different models are considered: The first is the ``light seed'' model (i.e., the popIII model), which posits that supermassive black holes originate from the remnants of Population III star evolution. The second is the ``heavy seed'' model (i.e., the Q3d model), which assumes that massive black holes with masses of $10^5 \ M_{\odot}$ had already formed in the early universe $(z\approx15-20)$, potentially due to galaxy collisions or other factors. This model accounts for the time delay effects between \ac{MBBH} mergers and galaxy mergers. The third model (i.e., the Q3nod model) is almost identical to the second but neglects the so-called delay effects. The semi-analytical calculations reveal that in the popIII model, LISA could detect approximately 660 \ac{MBBH} merger events over a five-year operational period. In the Q3d model, the total event rate is about 40, while in the Q3nod model, it is approximately 596. Therefore, future space-based detectors like LISA, TianQin and Taiji are expected to detect dozens to hundreds of \ac{MBBH} merger events, providing valuable observational samples for studying cosmology, particularly the evolution of the high-redshift universe.

\section{GW Sources as Cosmological Probes}
\label{sec3}

Similar to how \acp{SN Ia} can serve as standard candles in cosmology, for GW sources to act as ``standard sirens'', it is necessary to independently measure both the distance and redshift information of the GW event. Therefore, various methods have been developed to achieve these measurements. In this paper, we will briefly introduce the methods for measuring distance and redshift for different types of GW events, respectively.

\subsection{Determining the Distance of GW Sources} 
\label{sec3.1}

\subsubsection{Using the GW Waveform}

In 1986, Schutz first pointed out that by observing the GW waveform from a compact binary merger, one can independently measure the luminosity distance to the GW source \citep{1986Natur.323..310S}. The basic principle of this method is as follows: The amplitude of the GW depends on the chirp mass (a combination of the binary masses) of the source and its luminosity distance. Simultaneously, the masses can be precisely determined by measuring the phase of the GW signal. Therefore, by measuring both the amplitude and phase information of the source, one can obtain its luminosity distance. It is important to note that this distance measurement method successfully avoids the cosmic distance ladder, which is relied upon in general cosmological distance measurements, thereby circumventing various systematic errors associated with it. This is currently the primary GW source distance measurement method under discussion.

Mathematically, GWs are typically described by a tensor $h_{\mu\nu}$. In General Relativity, considering the transverse-traceless gauge, GWs contain two independent polarization components, $h_{+}$ and $h_{\times}$. For a GW detector, the observable is a linear combination of these two polarization components, i.e.,
\begin{equation}
h(t)=F_{+}(\theta,\phi,\psi) h_{+}(t)+F_{\times}(\theta,\phi,\psi) h_{\times} (t),
\end{equation}
where $F_{+}$ and $F_{\times}$ are the detector's response functions, which depend on the GW's polarization angle $\psi$ and the source's sky position angles $(\theta,\phi)$. For a network of multiple detectors, they also depend on each detector's location on Earth, orientation, opening angle between the arms, as well as the event's burst time and duration, etc. Therefore, in general, the response functions are also functions of time \citep{2018PhRvD..97f4031Z}.

Consider a merging compact binary system with luminosity distance $d_{\rm L}$, and binary component masses $m_1$ and $m_2$. The total mass is denoted as $M=m_1+m_2$, and the symmetric mass ratio as $\eta=m_1m_2/M^2$. We define the chirp mass as $\mathcal{M}_c=M\eta^{3/5}$. For sources at cosmological distances, the quantity entering the GW waveform is the so-called observed chirp mass, which is related to the physical chirp mass by: $\mathcal{M}_{c,{\rm obs}}=(1+z)\mathcal{M}_{c,{\rm phys}}$. Considering the GW amplitude to the lowest order approximation, during the inspiral phase, the two polarization components of the GW can be written as \citep{2011PhRvD..83b3005Z}
\begin{subequations}
 \begin{align}
 h_{+}(t)&= 2\mathcal{M}_c^{5/3} d_{\rm L}^{-1}(1+\cos^2\iota) \omega^{2/3}(t_0-t)  \cos[2\Phi(t_0-t;M,\eta)+\Phi_0],  \\
 h_{\times}(t)&= 4\mathcal{M}_c^{5/3} d_{\rm L}^{-1}\cos\iota~ \omega^{2/3}(t_0-t)  \sin[2\Phi(t_0-t;M,\eta)+\Phi_0],
 \end{align}
\end{subequations}
where $\iota$ is the inclination angle of the binary's orbital plane relative to the line of sight, $\omega(t_0-t)$ is the angular velocity of the equivalent one-body system orbiting the system's center of mass, and $\Phi(t_0-t;M,\eta)$ is the corresponding orbital phase. The constants $t_0$ and $\Phi_0$ are the time of merger and the phase angle at merger, respectively. Here, the effects of binary spins are not considered, and the orbit is assumed to be nearly circular. The phase angle $\Phi$ can be calculated using the post-Newtonian method, which has currently been computed up to the 3.5 post-Newtonian order. During the inspiral phase, the rate of change of the binary's orbital period is negligible. Therefore, the stationary phase approximation is generally used to perform a Fourier expansion of the GW waveform. Its Fourier component is
\begin{equation}\label{Hf}
H(f)=Af^{-\frac{7}{6}}\exp[i(2\pi f t_0-\pi/4+2\psi(f/2)-\varphi_{(2,0)})],
\end{equation}
where the Fourier amplitude is
\begin{equation}
A=\frac{1}{d_{\rm L}}\sqrt{F_+^2(1+\cos^2\iota)^2+F_{\times}^2\cos^2\iota}\sqrt{\frac{5\pi}{96}}\pi^{-\frac{7}{6}}\mathcal{M}_c^{\frac{5}{6}}.
\end{equation}
The functions $\psi$ and $\varphi_{(2,0)}$ are respectively
\begin{align}
 \psi(f)&= -\psi_0+\frac{3}{256\eta}\sum_{i=0}^{7}\psi_i (2\pi M f)^{i/3},  \\
 \varphi_{(2,0)}&= \tan^{-1}\left(-\frac{2\cos\iota ~F_{\times}}{(1+\cos^2\iota)F_+}\right). 
\end{align}
Note that the high-frequency cutoff for this approximation is generally chosen at $f_{\rm upper}=2f_{\rm ISCO}$, where the frequency of the innermost stable circular orbit (ISCO) is $f_{\rm ISCO}=1/(6^{3/2}2\pi M_{\rm obs})$.

From Equation (\ref{Hf}), we observe that the GW waveform $H(f)$ depends on nine independent parameters $(\mathcal{M}_c,\eta,d_{\rm L},\theta,\phi,\psi,\iota,t_0,\psi_0)$. Therefore, by analyzing the nine-parameter model of the GW, using methods such as the Fisher matrix or Monte Carlo techniques, constraints on these nine model parameters can be obtained. By marginalizing over the other parameters, constraints on the luminosity distance $d_{\rm L}$ can be derived. This is the fundamental reason why GW sources can serve as standard sirens. It should be noted here that there exist correlations among these nine model parameters, particularly a strong correlation between the distance parameter $d_{\rm L}$ and the inclination angle parameter $\iota$. Therefore, if other methods can determine the source's inclination angle in advance, it will significantly improve the constraint on the distance parameter $d_{\rm L}$ \citep{2014ApJ...795...43F,2017PhRvL.119r1102F, 2019PhRvX...9c1028C, 2024ApJ...974L..16S}. For example, if the gamma-ray burst produced by a \ac{BNS} merger can be observed, and since such bursts are generally believed to be concentrated within a small range of $\iota$, it not only determines the source's sky position $(\theta,\phi)$ but also the other two angles $(\iota,\psi)$, thereby greatly reducing the measurement error of $d_{\rm L}$ \citep{2017PhRvL.119r1102F, 2017ApJ...851L..36G, 2022PhRvD.106b3011W, 2023ApJ...943...13W}.

\subsubsection{Using Strong Gravitational Lensing Effects}
\label{lensGW}

Recently, \cite{2017NatCo...8.1148L} proposed a novel method for distance measurement to GW sources. The core idea is as follows. Assuming the GWs from a source undergo strong gravitational lensing (e.g., by a massive galaxy or galaxy cluster acting as the lens) on their way to the detector, multiple GW images can be produced, with time delays between them. These time delays consist of two effects: one geometric and the other the Shapiro time delay. Therefore, by measuring the arrival time differences between different images, one can infer the distance information of the GW source, providing a new method for distance measurement. Unlike other methods, the advantage of this approach is that the distance measurement does not rely on the amplitude measurement of the GW, thus avoiding uncertainties associated with system calibration. This method can be applied not only to compact binary merger events but also to other GW burst events, such as supernova explosions.

Specifically, according to strong gravitational lensing theory, the arrival time difference between different images can be calculated as:
\begin{equation}
\Delta t_{i,j}=\frac{D_{\Delta t}(1+z_d)}{c}\Delta \phi_{i,j},
\end{equation}
where $\Delta t_{i,j}$ is the arrival time difference between images $i$ and $j$, and $\Delta\phi_{i,j}=[(\tilde{\theta}_i-\beta)^2/2-\tilde{\psi}(\tilde{\theta}_i)-(\tilde{\theta}_j-\beta)^2/2+\tilde{\psi}(\tilde{\theta}_j)]$ is the difference in the Fermat potentials for images located at different positions $\tilde{\theta}_i$ and $\tilde{\theta}_j$, with $\beta$ representing the spatial position of the source. $\psi$ denotes the two-dimensional gravitational lensing potential, which depends on the critical density $\Sigma=c^2 D_{\rm s}/(4\pi G D_{\rm d}D_{\rm ds})$, where $D_{\rm s}$, $D_{\rm d}$, and $D_{\rm ds}$ are the angular diameter distances to the source at redshift $z_{\rm s}$, to the lens at redshift $z_{\rm d}$, and between the lens and source, respectively. In actual observations, for a given strongly lensed GW event, $\Delta t_{i,j}$, $z_{\rm s}$, $z_{\rm d}$, and $\Delta \phi_{i,j}$ can all be obtained through GW and electromagnetic counterpart observations. Therefore, the so-called time-delay distance $D_{\Delta t}$ can be derived from the above formula. Theoretically, this quantity is given by:
\begin{equation}
D_{\Delta t}=\frac{D_{\rm d}(z_{\rm d})D_{\rm s}(z_{\rm s})}{D_{\rm ds}(z_{\rm d},z_{\rm s})}.
\end{equation}
Thus, observing $D_{\Delta t}$ is essentially equivalent to observing a combination of three angular diameter distances, which in turn constrains cosmological evolution models. This is the core idea of the method. 
\cite{2017NatCo...8.1148L}
also estimated the event rate: considering third-generation GW detectors (e.g., ET), the annual event rate is expected to reach $50-100$ events, allowing for precise constraints on cosmological parameters such as the Hubble constant. It is estimated that observing about 10 such GW events with observed electromagnetic counterparts could constrain the Hubble constant to a level of $0.7\%$, which is already higher than the current constraints from traditional optical methods.

\subsection{Determining the Redshift of GW Source} 
\label{sec3.2}

To serve as standard sirens, GW sources must have their redshift information independently measured. However, from the two methods discussed above, the latter does not inherently contain a mechanism for independent redshift measurement. The former method relies primarily on the GW waveform, but the waveform from compact binaries does not directly depend on redshift $z$; its information is only contained within the observed chirp mass $\mathcal{M}_{c,\rm obs}=(1+z)\mathcal{M}_{c,\rm phys}$. Therefore, the redshift is completely degenerate with the source's intrinsic mass. It is difficult to obtain redshift information from GW observations alone. To solve this problem, various methods have been developed to extract the redshift information of GW sources. In this section, we will focus on introducing the following seven methods:

\subsubsection{Using EM Counterparts}

In general, the redshift information of a GW burst source needs to be obtained through observations of its electromagnetic counterpart. For example, during or after a \ac{BNS} merger or a NSBH merger event, a significant amount of electromagnetic radiation is produced almost simultaneously. If atomic spectral lines can be measured in this radiation (e.g., in the gamma-ray afterglow) via corresponding electromagnetic means, its redshift can be determined \citep{2007PhR...442..166N}. For instance, the GW event GW170817 was identified optically, leading to the discovery of its host galaxy NGC4993, from which its redshift (or equivalent recessional velocity) was measured \citep{2017Natur.551...85A}. 

Observations of GW170817 show that the electromagnetic counterparts of such GW bursts are very rich, covering the entire electromagnetic spectrum from gamma-rays to radio waves \citep[see][for a review]{2017ApJ...848L..12A}. Two seconds after the binary merger, Fermi/GBM and INTEGRAL detected the gamma-ray burst GRB170817A \citep{2017ApJ...848L..13A}. Subsequently, the electromagnetic counterpart SSS17a (also named DLT17ck or AT2017gfo) was discovered in optical, near-infrared, and ultraviolet bands between 0.47 and 18.5 days. 
An X-ray outburst was detected 9 days after the GW burst, and a radio outburst was found 16 days later \citep{2017Natur.551...71T}. 
Here, we primarily focus on the optical and near-infrared electromagnetic counterparts. When two neutron stars merge, approximately $(10^{-3}-10^{-2})\ M_{\odot}$ of neutron star material is ejected into space.
Within the ejected material, rapid neutron-capture nucleosynthesis, i.e., the $r$-process, occurs, producing radioactive heavy elements \citep{2017Natur.551...80K}. 
Subsequently, the decay of these radioactive elements generates optical and near-infrared radiation. This outburst process is commonly referred to as a kilonova (or macronova) \citep{2017Natur.551...75S}. Observations of SSS17a revealed that its absolute magnitude in the optical and near-infrared can reach -15 to -16 magnitudes (AB magnitude). The optical emission decays relatively quickly, lasting about two days, while the near-infrared emission persists longer. Since the optical and near-infrared radiation from kilonovae is approximately isotropic, in principle, for nearby events, it is possible to perform optical identification using telescopes and obtain their redshift information.

The final second-generation GW telescope network is expected to observe the gravitational radiation from the majority of \ac{BNS} mergers 
with
redshift $z<0.1$, with an angular resolution on the order of 10 square degrees. This corresponds well to the field of view of many next-generation optical telescopes. Here, we briefly discuss the Wide Field Survey Telescope (WFST) \citep{2023SCPMA..6609512W} as an example. WFST is a large-field survey telescope led by the University of Science and Technology of China and the Purple Mountain Observatory. With a 2.5-meter aperture and a large 3-degree diameter field of view (i.e., 7 square degrees), it is a dedicated astronomical telescope for large-scale, rapid imaging surveys, capable of achieving uniform, high-quality imaging across its 3-degree field. Its survey speed is 600 square degrees per hour, enabling a full survey of the northern sky every three days. According to the telescope design, short-exposure (30 seconds) images can reach a limiting magnitude of $r\sim23$, while long exposures can detect faint objects down to $r\sim25$. Considering that kilonovae have an absolute magnitude of -15 to -16, and WFST has a large 7-square-degree field of view, it is expected to capture the optical counterparts of the majority of \ac{BNS} mergers located in the northern sky with redshift $z<0.1$ \citep{2023ApJ...947...59L}. Recently, based on a target of opportunity (ToO) survey of GW candidate S250206dm, WFST has obtained the most stringent constraints on the electromagnetic counterpart of an NSBH merger event to date \citep{2026ApJ..1000L..20L}. In the near future observations, WFST will act as the most powerful survey detector in the northern hemisphere. Together with LSST, it can form a powerful detector network for the full-sky surveys, which will be essential for the kilonova detection during O4 and O5 \citep{2023ApJ...947...59L}. In addition to BNS and NSBH, the BBH events may also have the EM counterparts, if the mergers occurred in the gas-rich environment, {\it e.g.} in the accretion disk of \acp{AGN}. One typical example is the BBH event GW190521, a possible associated flare was discovered in \ac{AGN} J124942.29+344929.0 \citep{GW190521}. If confirmed, the redshifts of this kind of GW events can be fixed by the spectral observations of the host \acp{AGN}, and BBHs can become a new type of standard sirens \citep{HeLeia,Heleib,Heleic}. 

\begin{figure}[!htb]
\begin{center}
\includegraphics[scale=0.6]{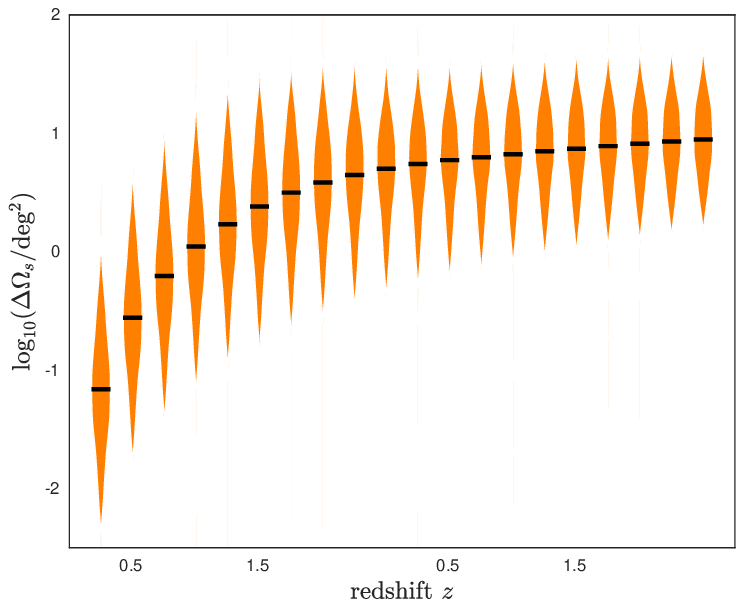}
\caption{Localization capability of a third-generation GW detector network (comprising three ETs located in the United States, Europe, and Australia) for \ac{BNS} merger events \citep{2018PhRvD..97f4031Z}. }
\label{a3}
\end{center}
\end{figure}

For third-generation GW detectors, their detection depth and spatial localization capabilities will be significantly improved (as shown in Figure \ref{a3}). For instance, ET is expected to detect the vast majority of \ac{BNS} merger events at redshifts $z<2$ with high confidence, leading to an estimated annual event rate reaching hundreds of thousands. For the nearby events, the localization capability of detector networks makes it possible to directly identify their host galaxies or galaxy groups, and therefore the redshifts can be determined \citep{Yu2024}. However, for the distant ones, the direct identification of their hosts through GW detector becomes impossible. Moreover, the corresponding kilonovae at $z>0.1$ are relatively difficult to observe, 
except for a (rare) population of binary neutron star mergers that produce magnetar remnants. In such cases, the heat from the magnetic winds driven by the remnant magnetars greatly boosts the kilonova luminosities, making them potentially observable at higher redshift \citep[e.g.,][]{2025MNRAS.544.3883C}.
Therefore, for such sources, the primary electromagnetic counterparts are generally considered to be short gamma-ray bursts and their afterglows \citep{2007PhR...442..166N}.  Their redshift information relies mainly on observations of atomic spectral lines in their optical afterglows. It is generally believed that short gamma-ray bursts are highly beamed, meaning only a small fraction of GW bursts will have observable shGRB counterparts. Additionally, factors such as the short duration of gamma-ray bursts and the challenges in measuring their optical afterglows contribute to the difficulty. In general discussions, a conservative estimate is often adopted, assuming that over a three-year observation period, approximately 1000 \ac{BNS} mergers will have identifiable electromagnetic counterparts and determined redshifts. It should be noted that space-based detectors like BBO and DECIGO will have significantly improved spatial resolution compared to ground-based detectors, reaching $1-100~{\rm arcsec}^2$, and are expected to directly identify host galaxies to determine redshifts \citep{2009PhRvD..80j4009C}. Consequently, the number of GW sources usable as standard sirens would greatly increase, reaching the order of millions.

Space-based LISA primarily observes \ac{MBBH} merger events, which are generally believed to be accompanied by intense electromagnetic outbursts, and thus are also expected to have rich electromagnetic counterparts. However, the key to successfully identifying electromagnetic counterparts and measuring redshifts lies in the GW detector having sufficiently sensitive spatial localization capabilities. If its resolution can reach around 10 square degrees, it would be comparable to the field of view of instruments like the Rubin Observatory Legacy Survey of Space and Time (LSST) \citep{2019ApJ...873..111I}, WFST \citep{2023SCPMA..6609512W}, the Extremely Large Telescope (ELT) \citep{2023ConPh..64...47P}, and SKA (Square Kilometre Array) \citep{2009IEEEP..97.1482D, 2015aska.confE..55M, 2015aska.confE.174B}, making electromagnetic identification, particularly of transient sources, feasible. In practical observations, there are two main scenarios for determining redshifts: First, if the optical counterpart is bright enough to be directly observed by optical telescopes like LSST, the redshift can be obtained through direct measurement. Second, if the electromagnetic counterpart appears as a radio jet or flare, it might be detected by SKA. Using the extremely high spatial angular resolution of radio telescopes to pinpoint the host galaxy, its redshift can then be obtained through spectroscopic or imaging observations of the galaxy with optical telescopes \citep{2016JCAP...04..002T}. 
\cite{2016PhRvD..93b4003K}
estimated the event rate for \ac{MBBH} mergers with determinable redshifts. 
They found that, considering a five-year LISA observation period, the number of such events is expected to be approximately $(1-450)$, with the specific number depending on the final design of LISA and the formation mechanisms of \acp{MBBH}.

\subsubsection{Using the Redshift Distribution of Host Galaxies or Galaxy Clusters}

For a large number of GW sources whose redshifts cannot be determined through optical identification (e.g., \ac{SBBH} mergers, high-redshift \ac{BNS} mergers, \ac{MBBH} mergers, etc.), their spatial positions can still be determined solely via GW observations. Within this angular region, there may exist several, or even hundreds, of galaxies or galaxy clusters. The redshift of each galaxy can be obtained through optical observations, thus analyzing the redshifts of these galaxies yields a redshift distribution function, which serves as a probability distribution for the GW source's redshift. Considering this redshift distribution, along with the distance constraints obtained from GW observations, such a GW source can also serve as a standard siren to constrain cosmology \citep{dark1,dark2,song,dark3,2022RAA....22a5020C}. 
\citet{2008PhRvD..77d3512M} and \citet{2011ApJ...732...82P}
found that by considering LISA observations and the redshift distribution of host galaxies or galaxy clusters, the Hubble constant could be constrained to a precision of $1\%$, or the dark energy equation of state parameter to a precision of $4-8\%$. 
Using a similar method for a network of second-generation ground-based GW detectors, 
\citet{2012PhRvD..86d3011D} found that if 50 GW events are detected, the Hubble constant could be determined to a precision of a few percent. 
In real observations, the constraining power on the Hubble constant from GW sources without electromagnetic counterparts is also susceptible to the completeness of surveyed galaxy catalogs, which can significantly weaken the constraints. When only candidate host galaxies are used to provide redshifts, the constraint on $H_0$ derived by LIGO/Virgo/KAGRA data from hundreds of GW events reaches only about 30\% precision \citep{2025arXiv250904348T, 2026arXiv260527227T}. 

Besides using candidate host galaxies within the localization error volume of GW sources to indicate their redshift distribution, at the technical aspect of data analysis, one can also treat the spatial distribution of the GW source population as a manifestation of the cosmic large-scale structure in luminosity distance space (assuming an astrophysical origin for the sources). 
By cross-correlating this with the traditional large-scale structure mapped by galaxy surveys in redshift space, the transformation between luminosity distance and redshift can be constrained, i.e., cosmological parameters can be constrained. 
The literatures \citep{2016PhRvD..93h3511O,2018arXiv181107136Z,2020ApJ...902...79B,2020PhRvR...2b3314C,2021PhRvD.103d3520M} have explored the prospects of using this GW-galaxy cross-correlation analysis to constrain cosmological parameters with future gravitational wave observations, and \citet{2024ApJ...975..189M} applied it to actual LIGO/Virgo/KAGRA data, demonstrating promising constraining power.

\subsubsection{Using the Redshift Distribution Function of Compact Object Binaries}

Recently, 
\cite{2019JCAP...04..033D}
proposed a novel method that
the estimation of cosmological parameters only requires considering the distance measurements of GW sources and their uncertainties. The redshift information of the sources is derived from the distribution function of GW source redshifts, which is obtained by considering the merger rates of compact binaries at different redshifts calculated from population synthesis models. Thus, for third-generation GW detectors like a single ET, even if source localization is not possible, they can still be used to constrain cosmological parameters. Furthermore, similar to the second method mentioned above, this approach can be applied to any type of GW source, not limited to \ac{BNS} mergers. Therefore, in the era of third-generation detectors, the annual event rate could reach $10^3-10^7$, with the event redshift range extending up to $z\sim17$.

Specifically, considering cosmological parameters $\vec{\Omega}=(H_0,\Omega_m)$, where $H_0$ is the Hubble constant and $\Omega_m$ is the matter density parameter, let $D_i$ represent the distance of the $i$-th GW source measured by the GW detector, and $z_i$ denotes its redshift. The probability distribution function for the cosmological parameters can be expressed as:
\begin{equation}
P(\vec{\Omega}|\vec{D})=\Pi_{i=1}^{n}\int _0^{z_{\max}}P(\vec{\Omega},z_i|D_i) dz_i,
\end{equation}
where,
\begin{equation}
P(\vec{\Omega},z_i|D_i)\propto P(D_i|\vec{\Omega},z_i)P(z_i|\vec{\Omega}).
\end{equation}
Here, $P(D_i|\vec{\Omega},z_i)$ is the likelihood function of the observational data, which can be obtained through theoretical calculation. $P(z_i|\vec{\Omega})$ is the distribution function for the source redshift, which requires a theoretical prior assumption and can be numerically calculated in practice using population synthesis models. Through numerical simulations, 
it is found that for the ET detector, if $10^5$ source distance measurements achieve a precision of $10\%$, the error on the Hubble constant can be constrained to the level of $1\%$. 

\subsubsection{Using GW Phase Corrections due to Tidal Effects}

The key reason why GW waveform measurements cannot directly yield redshift information is that redshift and source mass always appear in the form of the observed mass, both in amplitude and phase, making redshift and mass completely degenerate. However, for a compact binary merger event where one of the objects is not a black hole (i.e., it could be a neutron star, white dwarf, or even a main-sequence star), tidal disruption effects become significant when the binary separation is very small. 
This causes the object to deform, introducing a new quadrupole moment. This tidal effect modifies the phase of the GW waveform. Interestingly, this correction depends on the physical mass of the star, not the observed mass \citep{2010PhRvD..81l3016H}. Therefore, if this phase correction term can be observed, it can break the degeneracy between redshift and physical mass, allowing the redshift of the source to be determined. 
\cite{2012PhRvL.108i1101M}
first conducted a quantitative analysis of this redshift determination method, while in \cite{2017PhRvD..95d3502D}, this method was used for the first time in a preliminary study on constraining cosmological parameters. The drawback of this method is that the tidal phase correction depends on the neutron star equation of state, which is currently not well understood.

Considering a \ac{BNS} merger event, if tidal effects are ignored, the GW waveform in Fourier space can be written in the form of Equation (\ref{Hf}). If tidal effects are considered, and only the lowest-order approximation is taken, the phase in this expression gains an additional term:
\begin{align}
\psi_{\rm tidal} = \sum_{a=1,2}\frac{3\lambda_a}{128\eta} & \left[ -\frac{24}{\chi_a} \Big( 1+\frac{11\eta}{\chi_a} \Big) \frac{x^{5/2}}{M^5} -\frac{5}{28\chi_a} \Big(3179-919\chi_a-2286\chi_a^2+260\chi_a^3 \Big) \frac{x^{7/2}}{M^5}\right],
\end{align}
where the post-Newtonian parameter is $x=(\pi M f)^{2/3}$, $a=1,2$ represent the two neutron stars respectively, and $\chi_a=m_a/M$. The parameter $\lambda=(2/3)R_{\rm ns}^5 k_2$ describes the strength of the quadrupole moment induced by the external tidal field, which depends on the neutron star radius $R_{\rm ns}$ and the tidal Love number $k_2$ from the neutron star equation of state. Note that the masses $m_a$ and $M$ here refer to physical masses, not observed masses. Therefore, considering the tidal correction to the phase can break the mass-redshift degeneracy and allow the source redshift to be determined using GW observations alone. Although this effect is a fifth post-Newtonian order correction, because its coefficient is relatively large, about $O(R_{\rm ns}/M) \sim 5$ for neutron stars, the magnitude of the correction term can reach the level of general 3.5 or even 3 post-Newtonian order corrections. 
\cite{2012PhRvL.108i1101M}
found that for sources at redshift $z<1$, using this method, the ET detector could constrain the source redshift to a precision of $8-40\%$, while for sources at $1<z<4$, the redshift precision would be $9-65\%$. The main uncertainty of this method lies in the fact that the tidal deformation of neutron stars depends on their equation of state, which is currently unknown. This uncertainty greatly limits the applicability of the method. 
{If in the future we are able to precisely constrain the equation of state of neutron stars, then using this method, the third-generation GW detector network holds the potential to achieve constraints on the Hubble constant and the dark energy equation of state at the levels of 0.1\% and 1\% \citep{2023JCAP...08..070J}, respectively. 
}

Recently, we found that if low-redshift observations of GW electromagnetic counterparts can be combined, strict constraints on the neutron star equation of state can be achieved through joint observations of low-redshift GWs and their kilonova optical counterparts \citep{2020ApJS..250....6W}. 
Extending these constraints to high redshifts would increase the number of GW sources usable as standard sirens by more than three orders of magnitude. This could significantly improve the capability of the standard siren method to constrain cosmological parameters like dark energy and potentially become one of the primary methods for future dark energy exploration \citep{2020ApJS..250....6W}.

Using tidal effects to break the mass-redshift degeneracy can also be applied to the GW waveform of the ringdown phase after binary merger. This idea was recently discussed in detail in \cite{2014PhRvX...4d1004M}. The basic idea is as follows: when two neutron stars merge, they form a hyper-massive neutron star within a very short time. This neutron star undergoes bar-mode deformation for less than a second before collapsing into a black hole, simultaneously radiating GWs with a characteristic frequency band. This characteristic frequency depends on the total physical mass of the system. 
Through numerical simulations, 
they found this frequency can be approximated by a polynomial function of the physical mass, thus breaking the mass-redshift degeneracy. Simulating ET observational data revealed that for GW sources at redshift $z<0.04$, this method can constrain the source redshift to a precision of $10-20\%$.

\subsubsection{Using the Mass Distribution Function of Neutron Stars}

For \ac{BNS} merger systems, if the physical mass of the neutron stars is known in advance, the redshift of the GW source can be directly obtained from the difference between the observed mass measured from GWs and the physical mass. So, can current neutron star theory and observations provide definitive information about neutron star masses? In recent years, masses of many nearby neutron stars have been observed, revealing a multi-peaked distribution, reflecting differences in their formation mechanisms. However, for \ac{BNS} systems, observations indicate that the neutron star masses follow an approximately Gaussian distribution. 
For example, 
\cite{2013ApJ...778...66K}
found that in \ac{BNS} systems, the neutron star mass has a mean of $\sim 1.34\ M_{\odot}$ and a variance of $\sim 0.06\ M_{\odot}$, showing a peaked distribution structure. Furthermore, from a theoretical perspective, population synthesis theory also predicts that neutron star masses in \ac{BNS} systems exhibit a peaked distribution, with the peak around $1.3\ M_{\odot}$, which can be approximated by a Gaussian function. Considering cosmological redshift, the neutron star mass measured via GW waveform analysis will inevitably deviate from this Gaussian distribution, showing a clear distinction from the physical mass distribution. 
Assuming that the neutron star mass follows a certain Gaussian distribution, 
with its peak and variance as variables, through Bayesian analysis, \cite{2012PhRvD..85b3535T} found that cosmological parameters and the neutron star mass distribution function could be simultaneously constrained. This method also avoids the need for electromagnetic counterpart observations for \ac{BNS} mergers. The analysis found that for second-generation GW detectors, if 100 \ac{BNS} merger events can be observed, the Hubble constant could be constrained to a precision of $10\%$, while the error on the peak of the neutron star mass distribution could be constrained to $0.012\ M_{\odot}$, and the precision of the Gaussian distribution's standard deviation could reach $0.07\ M_{\odot}$. 
Considering the third-generation detector ET, 
\cite{2012PhRvD..86b3502T}
considered $10^5$ GW events and found that the capability to probe dark energy could reach the precision level of future optical methods. It should be noted that the application of this method relies on prior assumptions about the neutron star mass distribution. However, the electromagnetic and GW observation determined BNS mass distributions are dependent on different selection biases and different environments, while the distribution resulting from population synthesis model is model dependent as various not well understood physical processes are involved in the model \citep{2017ApJ...846..170T, 2020MNRAS.494.1587C, 2022MNRAS.509.1557C, 2025ApJ...980..181C}. 
With the discovery of more neutron stars and an increase in GW events, the traditional Gaussian distribution might undergo significant changes, introducing some uncertainty to the application of this method.

\subsubsection{Using the mass spectrum of Stellar-mass Black Holes}

For binary black hole systems, while no physical law confines their masses to a specific range, the mass distribution of \acp{SBBH} is not arbitrary if they are of astrophysical origin (i.e., remnants of stellar evolution) \citep{2012ApJ...749...91F, 2017ApJ...836..244W}. 
This constrained physical mass distribution can, to some extent, break the degeneracy between mass and redshift in GW observations, effectively constraining the redshift information of the GW source without relying on additional external observations \citep{2019ApJ...883L..42F}. 
Stellar population synthesis and evolutionary theories predict that the mass distribution of \acp{SBBH} approximately follows a power-law form from low to high masses, with a truncation around $\sim 45\,M_\odot$ \citep{2017ApJ...851L..25F, 2019ApJ...883L..42F, 2022ApJ...941L..39W}. 
The range from $\sim 45\,M_\odot$ to $\sim 120\,M_\odot$ corresponds to the theoretically predicted ``mass gap'' where stellar evolution is unlikely to form black holes directly, based on the pair-instability supernova mechanism \citep{2002ApJ...567..532H, 2016A&A...594A..97B, 2017ApJ...836..244W, 2024PhRvL.133e1401L, 2025arXiv251022698W, 2026arXiv260511474L}. 

The population properties of \ac{SBBH} merger events detected by the LIGO/Virgo/KAGRA detector network have demonstrated that the black hole mass function is broadly consistent with a power-law distribution, featuring a global peak around $\sim 10\,M_\odot$ and providing preliminary evidence for the existence of the black hole mass gap \citep{2019ApJ...882L..24A}. 
However, the observed mass function also exhibits a localized peak around $\sim 35\,M_\odot$, which may be caused by hierarchical mergers of black holes \citep{2023PhRvX..13a1048A, 2025arXiv250818083T}. 
These characteristic features of the black hole mass function can effectively constrain the underlying physical mass distribution of the observed \ac{SBBH} \ac{GW} sources \citep{2019ApJ...883L..42F, 2024ApJ...976..153L, 2025ApJ...980...85M}. 
Consequently, the GW source's redshift can be inferred from the observed detector-frame mass $M_{\mathrm{obs}}$ and the physical mass $M_{\mathrm{phys}}$ via the relation $z = M_{\mathrm{obs}} / M_{\mathrm{phys}} - 1$.   
Methods that utilize the intrinsic mass distribution function of compact objects (including black holes and neutron stars) to directly extract redshift information of GW sources are collectively termed ``spectral sirens'' \citep{2022PhRvL.129f1102E}. 
Currently, the LIGO/Virgo/KAGRA collaboration has formally adopted the spectral siren method as one of the primary techniques for measuring the Hubble constant, demonstrating its potential for providing meaningful constraints \citep{2023ApJ...949...76A}. 
However, limited by the current total number of GW events and the precision of source parameter estimation, the constraints on the Hubble constant from the spectral siren method alone remain relatively weak \citep{2025arXiv250904348T, 2025arXiv250903607M}. 
It is therefore commonly used in conjunction with surveyed galaxy catalogs \citep{2023PhRvD.108d2002M, 2023JCAP...12..023G, 2025arXiv250904348T}. 

The reliability of extracting redshift information via the spectral siren method depends critically on the accuracy of the black hole mass distribution function model \citep{2021PhRvD.104f2009M, 2022PhRvL.129f1102E}. 
Present constraints on this model from LIGO/Virgo/KAGRA observations still carry significant uncertainties \citep{2025arXiv250818083T}. 
Furthermore, the formation of stellar-mass black holes is highly sensitive to the metallicity of their progenitor stars, implying that the \ac{SBBH} mass spectrum is likely cosmologically evolving \citep{2017MNRAS.470.4739S, 2022MNRAS.515.5495M}. 
Various features of the mass distribution function, such as the peaks around $\sim 10\,M_\odot$ and $\sim 35\,M_\odot$, as well as the $\sim 45 - 130\,M_\odot$ mass gap, could be functions of redshift. 
Fortunately, the greatest advantage of the spectral siren method lies in its ``self-calibrating'' property: 
as the sample of detected \ac{SBBH} merger events grows, the model for the \ac{SBBH} mass spectrum can be continuously refined \citep{2019ApJ...882L..24A, 2023PhRvX..13a1048A, 2025arXiv250818083T}. 
For third-generation ground-based GW detectors (e.g., ET and CE), millions of compact object binary merger events are expected to be detected \citep{2011PhRvD..83b3005Z,2018PhRvD..97f4031Z, 2021arXiv210909882E, 2026JCAP...03..081A}. 
Using the merger GW event sample detected within just one month, the reconstruction precision of the Hubble parameter $H(z)$ over the redshift range $z=0-3$ could significantly surpass the precision provided by the entire mission of surveys like DESI, enabling cosmological inference at the sub-percent level \citep{2022PhRvL.129f1102E, 2022PhRvD.105b3523L}. 

\subsubsection{Using GW Phase Corrections from Cosmic Evolution}

For the gravitational radiation from compact binary mergers, as mentioned earlier, the Fourier component of the GW can be expressed in the form of Equation (\ref{Hf}). This approximation does not include higher-order corrections due to the accelerated expansion of the universe. In fact, when calculating the propagation equations of photons or gravitons in the universe within the standard cosmological model and considering terms up to second order, the relationship between the time interval $\Delta t_{e}$ of the same event and the observed time interval $\Delta t_o$ is given by \citep{2001PhRvL..87v1103S}
\begin{equation}
\Delta t_o=\Delta T+X(z)\Delta T,
\end{equation}
where $z$ is the cosmological redshift at which the event occurs, $\Delta T=(1+z)\Delta t_e$, and
\begin{equation}
X(z)=\frac{1}{2}\left(H_0-\frac{H(z)}{1+z}\right),
\end{equation}
here $H_0$ is the Hubble constant, and $H(z)$ is the Hubble parameter at redshift $z$. It can be seen that the functional relation $X(z)$ directly depends on the cosmological model.

If this effect is taken into account, the phase in Equation (\ref{Hf}) will acquire a new correction term, known as the GW phase modulation due to cosmic acceleration \citep{2001PhRvL..87v1103S}
\begin{equation}\label{Psi_acc}
\psi_{\rm acc}=-\frac{25}{32768}X(z)M(\pi M f)^{-13/3}.
\end{equation}
Note that the mass $M$ here is still the observed total mass. Therefore, for a GW event from a binary merger, in principle, by observing the GW waveform, one can independently obtain the source's luminosity distance $d_{\rm L}$ and the function $X(z)$. Thus, the $d_{\rm L}-X$ relation, rather than the $d_{\rm L}-z$ relation, can be used to constrain cosmological models.

From expression of Equation (\ref{Psi_acc}), it can be seen that this effect belongs to the 4th post-Newtonian order correction, making it relatively difficult to observe. It can serve as a good cosmological probe only when the signal-to-noise ratio is high and the number of GW bursts is sufficiently large. Therefore, it is generally believed that this method is not suitable for ground-based laser interferometers (such as LIGO) or space-based laser interferometers like LISA.
\citet{2001PhRvL..87v1103S} and \citet{2012PhRvD..85d4047N} investigated the cases for space-based laser interferometers such as BBO and DECIGO,
and they found that the dark energy equation of state parameter can be constrained to a precision of $10\%$ for $10^6$ binary star merger GW events.

\section{Constraints on Cosmological Parameters}
\label{sec4}

This section introduces the capability of GW sources from compact binary mergers as ``standard sirens'' to constrain cosmological parameters. We separately discuss the constraining capability on the Hubble constant and dark energy for ground-based laser interferometer GW observatories (including second and third-generation detectors) and space-based laser interferometer GW detectors (including LISA, TianQin, Taiji, BBO, DECIGO, etc.), and compare them with traditional electromagnetic methods.

\subsection{Ground-based laser interferometer networks} 
\label{sec4.1}

\subsubsection{The Era of Second-generation Ground-based GW Detectors}

Second-generation ground-based GW detectors mainly include the currently operating two Advanced LIGO, Advanced Virgo and KAGRA detectors, as well as the future LIGO-India, forming a network of five detectors. 
For GW events from \ac{BNS} mergers, these detectors can detect nearby sources with redshifts $z\lesssim 0.1$. Since \ac{BNS} mergers and NSBH mergers are accompanied by rich electromagnetic radiation, and for these nearby sources, their electromagnetic (especially in the visible band) radiation can be fully observed by current and future telescopes, they can serve as ``standard sirens'' to study the cosmic expansion history. Within this redshift range, Hubble's law, which describes the relationship between cosmological redshift and luminosity distance, approximately holds. 
Therefore, observations of these GW sources provide a new avenue for measuring the Hubble constant. It should be noted that current general methods for measuring the Hubble constant are roughly divided into two categories: One uses nearby \acp{SN Ia}. For example, analysis of supernova observational data from SH0ES yields a constraint on the Hubble constant of $H_0=73.04\pm 1.04 \ {\rm km} \cdot{\rm s}^{-1}\cdot{\rm Mpc}^{-1}$ \citep{2022ApJ...934L...7R}.  
The characteristic of this method is that the measurement does not depend on the choice of other cosmological parameters but directly fits Hubble's law. However, it may be affected by various systematic errors \citep{2021CQGra..38o3001D, 2022NewAR..9501659P}. 
The other uses the temperature and polarization anisotropy power spectra of the Cosmic Microwave Background (CMB) to jointly constrain cosmological parameters, including the Hubble constant. For instance, recent fitting of Planck satellite observational data gives a Hubble constant of $H_0=67.4\pm 0.5  \ {\rm km} \cdot{\rm s}^{-1}\cdot{\rm Mpc}^{-1}$ \citep{2020A&A...641A...6P}. This method has relatively fewer systematic errors, but its drawback is that the constraint on the Hubble constant depends on the choice of the cosmological model. 
Currently, the results from the two methods show a discrepancy around $5\sigma$, which is one of the challenges in modern cosmology \citep{2017NatAs...1E.169F, 2021CQGra..38o3001D, 2022NewAR..9501659P}. 
There are two possible reasons for this inconsistency: First, there might be unknown systematic errors in at least one of the two measurement methods leading to significant deviations in the results. Second, it could be due to unknown new physics, such as evolving dark energy 
or sterile neutrinos. 
Therefore, GW observations of the Hubble constant may provide a new way to resolve this inconsistency in the future \citep{2019PhRvL.122f1105F, 2020ApJ...905L..28B, 2023PhRvD.107l3519C}.

\begin{figure}[!htb]
\begin{center}
\includegraphics[scale=0.5]{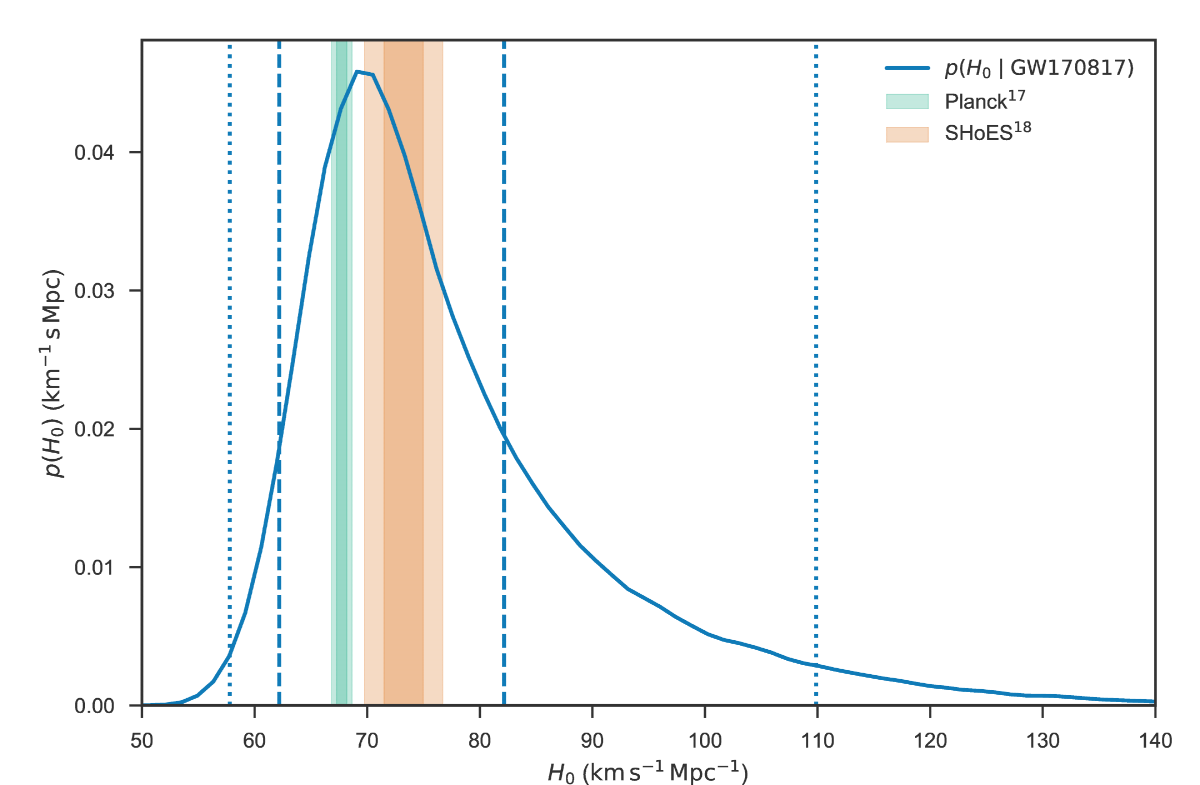}
\caption{Comparison of the Hubble constant constraint derived from GW170817 with results from the CMB and supernovae \citep{2017Natur.551...85A}. }
\label{a4}
\end{center}
\end{figure}

Now, let us use the GW170817 event as an example to introduce the capability of second-generation GW detectors to probe the Hubble constant \citep{2017Natur.551...85A}. As mentioned earlier, Hubble's law holds in the nearby universe, meaning the distance $d_{\rm L}$ and the Hubble flow velocity (or recession velocity) $v_{H}$ satisfy a linear relationship:
\begin{equation}
v_{H}=H_0 d_{\rm L},
\end{equation}
where the proportionality coefficient $H_0$ is the Hubble constant. Note that the relationship between the recession velocity $v_{H}$ and the cosmological redshift is $z=v_{H}/c$. Therefore, as long as $v_{H}$ and $d_{\rm L}$ are measured, constraints on the Hubble constant can be obtained. For the GW170817 GW event, its recession velocity was obtained through measurements of its host galaxy, NGC4993. Shortly after the GW burst, its host galaxy was identified through optical observations. The galaxy's velocity relative to the CMB rest frame is $3327\pm 72  \ {\rm km} \cdot{\rm s}^{-1}$. 
Considering the influence of the galaxy's peculiar velocity, the resulting galaxy recession velocity is $v_H=3017\pm 166 \ {\rm km} \cdot{\rm s}^{-1}$, equivalent to a cosmological redshift of $z=0.01006\pm 0.00055$ \citep[also see][for an additional virial
velocity correction]{2021A&A...646A..65M}. 
The measurement of the luminosity distance $d_{\rm L}$ comes from analyzing the GW data. In the template fitting, besides $d_{\rm L}$, there are multiple model parameters (such as neutron star masses, initial phase, merger time, system inclination, etc.). The distance constraint obtained by marginalizing over other parameters is $d_{\rm L}=43.8^{+2.9}_{-6.9}$ Mpc. This error includes statistical measurement errors, instrumental systematic errors, and the coupling between distance and inclination. 
Combining the two yields a constraint on the Hubble constant $H_0=70.0^{+12.0}_{-8.0} \ {\rm km} \cdot{\rm s}^{-1}\cdot{\rm Mpc}^{-1}$. As shown in Figure \ref{a4}, this result is consistent with current CMB and SN Ia measurements, initially demonstrating the feasibility of using GW sources as standard sirens for cosmology studies. However, due to the limited number of GW events, the constraint on the Hubble constant remains relatively weak and currently cannot match the precision of traditional methods.

In a recent article, \citet{2018Natur.562..545C}
systematically analyzed the constraining abilities of second-generation GW detectors on the Hubble constant. The study considered both \ac{BNS} and \ac{SBBH} mergers as GW events. To measure the redshift of the sources, the authors explored two approaches: one is the electromagnetic counterpart method, similar to GW170817; the other is the use of the redshift distribution of host galaxies or galaxy clusters (as described in the previous section). 
The analysis reveals that if approximately 10, 60, or 200 \ac{BNS} merger events along with their electromagnetic counterparts are observed, the Hubble constant can be constrained to an accuracy of 4\%, 2\%, or 1\%, respectively. Considering the event rate, the authors suggest that if second-generation detectors achieve their design sensitivity and operate successfully for two years, it may be possible to constrain the Hubble constant to an accuracy of 1\%, which meets or even exceeds the current precision levels achieved by traditional methods. Now, this prediction is challenged due to the little discovery of BNS during O3 and O4 \citep{2023PhRvX..13d1039A, 2025arXiv250818082T}, and we place the hope in the future O5 and the third-generation era.

\subsubsection{The Era of Third-generation Ground-based GW Detectors}

\begin{figure}[!htb]
\begin{center}
\includegraphics[scale=0.5]{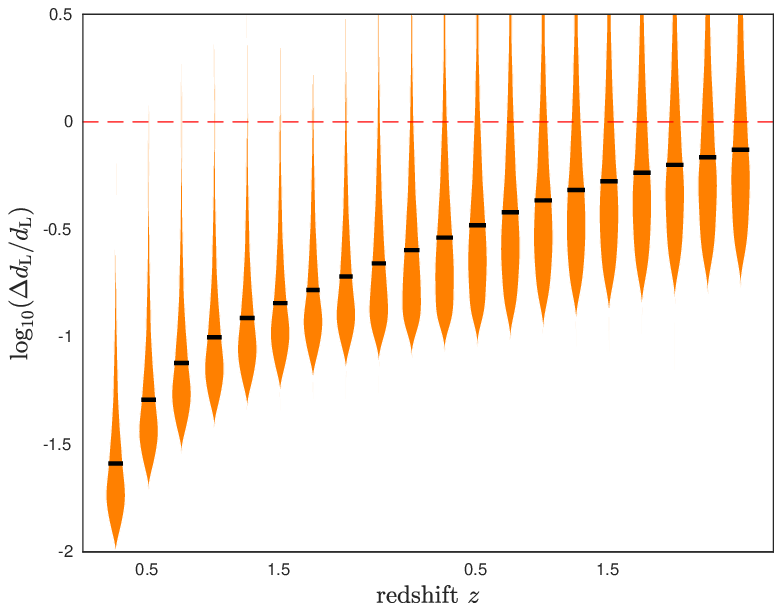}
\caption{Distance measurement capability of a third-generation GW detector network (comprising three ETs located in the United States, Europe, and Australia) for \ac{BNS} merger events \citep{2018PhRvD..97f4031Z}.}
\label{a5}
\end{center}
\end{figure}

Compared to the second generation, third-generation GW detector networks will significantly improve both the localization capability for nearby sources and the precision of distance measurements. The former facilitates the identification of electromagnetic counterparts, while the latter directly enhances the constraints on the Hubble constant. 
\cite{2018PhRvD..97f4031Z}
discussed in detail the constraining abilities of various possible third-generation detector networks on \ac{BNS} mergers and NSBH mergers. 
For example, considering a network of three ET located in the United States, Europe, and Australia, the distance measurement for a \ac{BNS} merger event at $z=0.1$ can achieve an accuracy of $10\%$–$1\%$ (see Figure \ref{a5}). If 100 such events with electromagnetic counterparts are observed at this redshift, the Hubble constant can be constrained to an accuracy of $0.3\%$.

\begin{figure}[!htb]
\begin{center}
\includegraphics[scale=1.0]{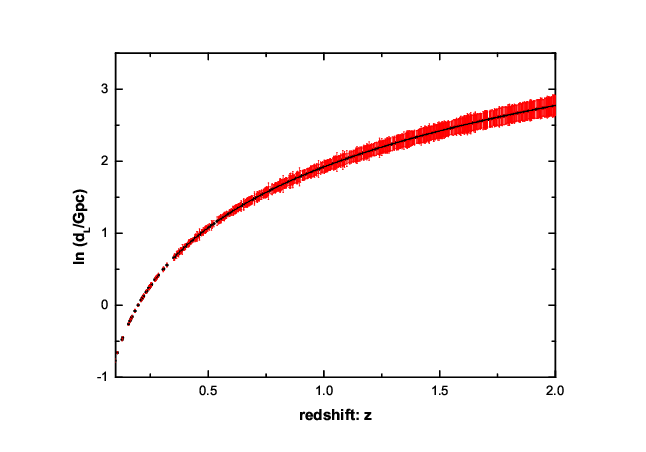}
\caption{Luminosity distances and their error distributions for 1000 \ac{BNS} merger GWs obtained from numerical simulations, where a third-generation detector network consisting of one ET and one CE is considered.}
\label{a6}
\end{center}
\end{figure}

More importantly, third-generation detectors can detect GW bursts at higher redshifts. For instance, for ET, the distance measurement accuracy for most events at $z<1$ will be better than $10\%$, and the two-dimensional sky localization will be smaller than 10 square degrees (see Figure \ref{a3}). Therefore, observations of these sources and their electromagnetic counterparts will enable the study of high-redshift cosmology, in addition to constraining the Hubble constant, they can be used to investigate the properties of dark energy, which is one of the most important topics in current cosmological research. 
As mentioned earlier, for GW sources at $z>0.1$, the observable electromagnetic counterparts are expected to be primarily beamed $\gamma$-ray bursts and their afterglows. Consequently, only a very small fraction of GW sources will have identifiable electromagnetic counterparts and determined redshifts \citep{2018ApJ...857..128J, 2021ApJ...916...54Y}. However, once the electromagnetic counterpart is identified for a given source, its spatial position, orbital inclination, and polarization angle can be determined, thereby greatly improving the distance measurement capability from GW observations. 
Note that for these higher-redshift sources, distance errors arise not only from instrumental noise but also from weak gravitational lensing, which is non-negligible \citep{2006PhRvD..73b3523B, 2010PhRvD..81l4046H, 2018ARA&A..56..393M} . 
For a source at $z=1$, weak lensing introduces approximately a $5\%$ error in the distance measurement \citep{2010CQGra..27u5006S}. In Figure \ref{a6}, we simulated the luminosity distances and their error distributions for 1000 \ac{BNS} merger GW sources with redshifts $z\in(0.1,2.0)$. The third-generation detector network considered here includes one CE and one ET, and we assumed the GW sources are uniformly distributed in three-dimensional cosmological space.

Next, we discuss the constraints on cosmological parameters from these sources. Consider a Friedmann-Lema\^itre-Robertson-Walker universe containing matter and dark energy, the dark energy equation of state parameter is parameterized in the following form \citep[CPL model;][]{2001IJMPD..10..213C, 2003PhRvL..90i1301L}:
\begin{equation}
w(z)=w_0+w_a\frac{z}{1+z},
\end{equation}
where $w_0$ is the present value of the equation of state parameter, and $w_a$ describes its evolution with redshift. For the cosmological constant model, we have $w_0=-1$ and $w_a=0$. This model includes five parameters: $w_0$, $w_a$, $H_0$, $\Omega_m$, and $\Omega_k$, where the latter two are the density parameters for dust matter and spatial curvature, respectively. For a given redshift $z$, the luminosity distance $d_{\rm L}$ is a function of these five parameters, i.e.:
\begin{equation}
d_{\rm L}(z)=d_{\rm L}(w_0,w_a,H_0,\Omega_m,\Omega_k;z).
\end{equation}
In principle, observations of distances and redshifts from 1000 standard sirens can simultaneously constrain these five cosmological parameters. However, in practice, \cite{2011PhRvD..83b3005Z} found that relying solely on GW and electromagnetic counterpart observations cannot strongly constrain any single parameter. This is due to a strong degeneracy between the background cosmological parameters ($H_0,\Omega_m,\Omega_k$) and the dark energy parameters ($w_0,w_a$), which must be broken by other observations. It is worth noting that this issue similarly exists for other cosmological probes, such as SN Ia and BAO. 
In general cosmological studies, CMB data are typically considered to break this degeneracy. 
\cite{2011PhRvD..83b3005Z} found that jointly constraining cosmological parameters with CMB and GW data is approximately equivalent to using CMB data to determine the background cosmological parameters ($H_0,\Omega_m,\Omega_k$), while using GW data alone to constrain the dark energy parameters ($w_0,w_a$). Therefore, in our data simulations, we only need to consider using GW data to constrain the dark energy parameters ($w_0,w_a$).

\begin{figure}[!htb]
\begin{center}
\includegraphics[scale=0.6]{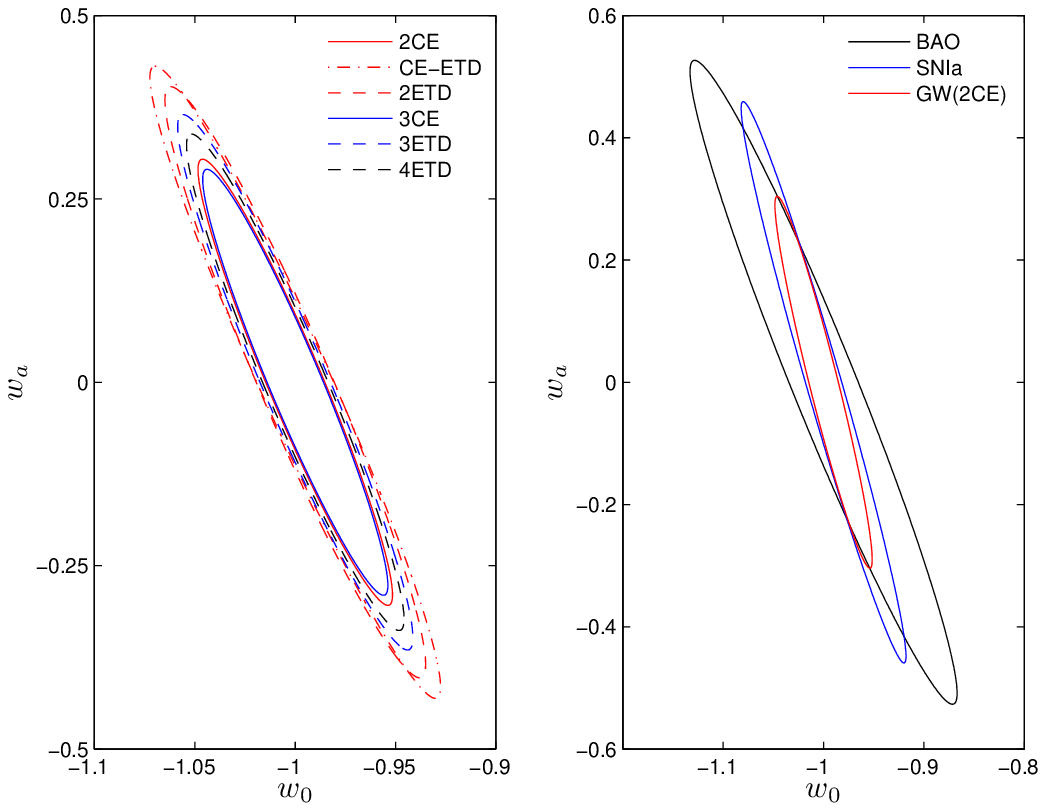}
\caption{\emph{Left panel}: Constraints on dark energy parameters from various third-generation GW detectors. \emph{Right panel}: Comparison of the constraining ability on dark energy between GW methods and other traditional methods (SN Ia and BAO). Note that for all GW detectors, we assume 1000 GW sources (see \citet{2018PhRvD..97f4031Z} for details).}
\label{a7}
\end{center}
\end{figure}

\cite{2018PhRvD..97f4031Z}
considered the constraining abilities of various third-generation GW detector networks on dark energy parameters. For example, considering a detector network comprising ET and CE, 1000 \ac{BNS} merger events yield the following constraints:
\begin{equation}
\Delta w_0=0.048,~~\Delta w_a=0.28.
\end{equation}
In Figure \ref{a7}, we present the constraining abilities on dark energy parameters for different third-generation detector networks and compare it with the projected capabilities of future SN Ia and BAO methods. From the figure, we find that the future GW method's constraining ability on dark energy can match or even surpass that of traditional optical methods. 
\cite{2017PhRvD..95d4024C} employed a dark energy parameter reconstruction method to study the constraining ability of GW standard sirens on dark energy, reached similar conclusions.

{Although the capability of GW siren method is quite optimistic in the third-generation era, it is also important to emphasize that the future practical application faces many challenges, including the accurate localization of GW events, the complete galaxy catalogs to identify the host galaxies, the reliable electromagnetic identification, the well-controlled selection effects and so on. Therefore, a complete, powerful, multi-band integrated collaborative observation network covering the ground-based and space-based telescopes is very important in the future.}

\subsection{Space-based Laser Interferometer} 
\label{sec4.1}

\subsubsection{The LISA Era}

Unlike ground-based GW detectors, space-based detectors such as LISA, primarily target the mergers of \acp{MBBH}. Consequently, utilizing these GW sources as standard sirens presents several distinctive features \citep{2023LRR....26....2A}: 
(1) The sources can extend to relatively high redshifts, reaching up to $z\sim 7$ or even higher, thus offering the potential to study cosmic evolution at high redshifts \citep{2016PhRvD..93b4003K,2016JCAP...04..002T, 2025CQGra..42s5002S}; 
(2) Compared to ground-based detectors, the number of such sources is relatively small, i.e. only a few tens, especially for events at redshifts $z<2$. 
Relying solely on these GW detections would provide very limited constraints on cosmological parameters (particularly dark energy parameters), necessitating joint analysis with other cosmological probes \citep{2016JCAP...04..002T}; 
(3) Due to the large masses of these sources, the influence of higher-order post-Newtonian terms is significant. Incorporating these correction terms can substantially improve the precision of GW parameter estimation \citep{2007PhRvD..76j4016A}; 
(4) Since the signal-to-noise ratio of the GW signals is often high, the final distance measurement error for the source is typically dominated by weak lensing rather than instrumental noise. Therefore, reconstructing the cosmic density field to mitigate the effects of weak lensing is crucial \citep{2010PhRvD..81l4031V, 2010PhRvD..81l4046H, 2026MNRAS.547ag353V}.

\cite{2016JCAP...04..002T}
discussed in detail the constraining ability on cosmological parameters for various design options of LISA project, considering the influence of different theoretical models for massive black hole formation. 
Here, we take one of the more ideal design concepts as an example: assuming an arm length of $5\times 10^{9}$ meters; low-frequency noise reaching the LISA Pathfinder level; a configuration with six links, equivalent to two independent Michelson interferometers; and a mission lifetime of five years. 
This concept is named N2A5M5L6 \citep{2017arXiv170200786A}. For a given GW source, the error in the luminosity distance arises from three aspects: measurement error due to instrumental noise, error from weak gravitational lensing, and error from the peculiar velocity of the host galaxy. 
For redshift measurement, the authors assumed that for sources with a signal-to-noise ratio greater than 8 and a two-dimensional sky localization better than 10 square degrees, their electromagnetic counterparts could be identified and redshifts determined using telescopes like LSST or SKA, with the potential redshift measurement error folded into the luminosity distance error. 
Based on three different models for \ac{MBBH} formation---popIII, Q3d, and Q3nod---the theoretically predicted numbers of detectable sources are 42, 15, and 32, respectively. 
If partial mitigation of the lensing effect and the contribution from the merger and post-merger ringdown phases are considered (referred to as the ``ideal case'' in the text), the source numbers increase to 46, 31, and 50, respectively.

Using Fisher matrix analysis, the constraints on cosmological parameters are as follows \citep[see][for a review]{2023LRR....26....5A}: If only the Hubble constant is constrained while other cosmological parameters are held fixed, the results are $\Delta h=0.00712$ (popIII), $0.00996$ (Q3d), $0.00531$ (Q3nod), where $h=H_0/(100 \ {\rm km} \cdot{\rm s}^{-1}\cdot{\rm Mpc}^{-1})$. 
In the ideal case, these constraints improve slightly to $\Delta h=0.00412$ (popIII), $0.00446$ (Q3d), $0.00307$ (Q3nod). On the other hand, if other cosmological parameters are fixed and only the dark energy parameters are constrained, the results are $\Delta w_0=0.253,\Delta w_a=1.32$ (popIII), $\Delta w_0=0.584,\Delta w_a=2.78$ (Q3d), $\Delta w_0=0.176,\Delta w_a=1.00$ (Q3nod). In the ideal case, these become $\Delta w_0=0.149,\Delta w_a=0.798$ (popIII), $\Delta w_0=0.241,\Delta w_a=1.14$ (Q3d), $\Delta w_0=0.101,\Delta w_a=0.544$ (Q3nod).
It is evident that, compared to third-generation ground-based GW detectors, LISA provides weaker constraints on dark energy. 
\citet{2017JCAP...05..031C}
found that LISA data could be used to reconstruct the redshift evolution of the interaction between dark matter and dark energy, allowing for relatively good constraints on the interaction within the redshift range $z\in(1,3)$. 
Therefore, LISA data provides a reliable avenue for studying high-redshift cosmology.

\subsubsection{The TianQin and Taiji Projects of China}

The advancement of China's space technology and the scientific potential of GW astronomy have led to the proposal of two Chinese space-based GW detection missions, 
TianQin \citep{2016CQGra..33c5010L, 2025CQGra..42q3001L} and Taiji \citep{2017NSRev...4..685H, 2022ChJSS..42..536L}, designed to operate alongside LISA. 
Both TianQin and Taiji employ a triangular constellation of three spacecraft, similar to LISA, forming two equivalent laser interferometers with their three arms. The key differences lie in their orbits and arm lengths. TianQin is designed for a geocentric orbit with an arm length of $\sim \sqrt{3} \times 10^8 ~\!{\rm meters}$, resulting in a slightly higher optimal sensitivity frequency band compared to LISA. Taiji, like LISA, is designed for a heliocentric orbit, but with its constellation center leading Earth's orbit by about 20 degrees. It features a longer arm length of $\sim 3 \times 10^9 ~\!{\rm meters}$, giving it a sensitivity band nearly identical to LISA's.

For probing the cosmic expansion history, TianQin and Taiji possess scientific potential comparable to LISA \citep{2020IJMPA..3550075R, 2025RPPh...88e6901L}. 
Both missions have strong detection capabilities for three primary types of cosmological-distance GW sources: MBBH mergers, EMRIs, and SBBH inspirals \citep{2022NSRev...9B..54W, 2026LRR....29....1L}. 
Taking the detection of MBBH mergers as an example, based on three population models (popIII, Q3d, and Q3nod), TianQin is projected to detect approximately 8, 4, and 26 events per year, respectively, with redshifts below $z \sim 3$ \citep{2019PhRvD.100d3003W}. 
Taiji's projected detection rates for MBBHs are similar to or somewhat exceed those of LISA.
Under the ideal assumption of observable electromagnetic counterparts, TianQin's potential for constraining cosmological parameters using MBBH mergers is as follows \citep{2022PhRvR...4a3247Z}:
(1) Simultaneously constraining the three parameters of the $\Lambda$CDM model ($H_0$, $\Omega_m$, $\Omega_\Lambda$), 
the expected precision on $H_0$ could reach approximately 4.3\% (popIII), 6.2\% (Q3d), and 1.9\% (Q3nod); 
on $\Omega_m$, approximately 20\% (popIII), 27\% (Q3d), and 7\% (Q3nod); and 
on $\Omega_\Lambda$, approximately 27\% (popIII), 28\% (Q3d), and 16\% (Q3nod).
(2) Simultaneously constraining the two parameters of the CPL 
dark energy model ($w_0$, $w_a$), while fixing $H_0$, $\Omega_m$, and $\Omega_\Lambda$, 
the expected precision on $w_0$ could reach approximately 12\% (popIII), 14\% (Q3d), and 9\% (Q3nod); and 
on $w_a$, approximately 66\% (popIII), 67\% (Q3d), and 49\% (Q3nod). 
Due to its lower optimal frequency band, Taiji has a stronger expected detection capability for low-frequency sources like MBBH mergers compared to TianQin. 
In the most optimistic scenarios, Taiji could achieve a measurement precision for $H_0$ approaching the 1\% level and a constraint on the CPL dark energy equation of state parameter $w_0$ better than 10\% \citep{2020SciBu..65.1340Z, 2022NSRev...9B..54W}.

In summary, while the scientific goals of TianQin, Taiji, and LISA have their respective emphases, there is significant overlap in their primary objectives \citep{2021PhRvD.104b4012W, 2024SCPMA..6759511T}. 
Joint observations by multiple detectors \citep{2020NatAs...4..108R, 2021NatAs...5..881G, 2024FunRe...4.1072C} can substantially enhance the constraints on cosmic expansion history. 
For instance, using MBBH mergers, combined observations by TianQin and LISA are expected to reduce the constraint errors on $H_0$ and $w_0$ by approximately 30\%--70\% compared to either mission alone \citep{2022PhRvR...4a3247Z}. 
The improvement from combined Taiji and LISA observations is projected to be even more significant \citep{2022NSRev...9B..54W, 2024SCPMA..6720412J}. 
{However, overall, due to the significant uncertainties in the astrophysical population properties of GW sources, the ability of future TianQin, Taiji, and LISA detectors to probe the cosmic expansion history remains considerably uncertain. 
One promising approach to enhance the scientific return of space-based GW data is to perform joint analyses with other high-redshift cosmological probes, such as the CMB and BAO \citep{2020JCAP...11..012W, 2022SCPMA..6510411W, 2024SCPMA..6720412J}, and the neutral hydrogen 21-cm intensity mapping \citep{2021PhRvD.104j3507J}. 
The constrained ``$D_L - z$'' relation at relatively low redshifts provided by space-based GW detectors can effectively anchor the cosmological constraints derived from these high-redshift data, extending them to lower redshifts with improved precision. This will help to test, in a manner independent of the cosmic distance ladder, the crises facing the standard cosmological model. 
}

\subsubsection{The Post-LISA Era}

Among the second-generation space-based GW detectors currently under discussion, two prominent examples are BBO \citep{2006CQGra..23.4887H} and DECIGO \citep{2006CQGra..23S.125K}. 
Taking BBO as an example, its sensitive frequency band is $(0.03-3)$ Hz. Its primary scientific goal is the detection of primordial GWs from the early universe in this band. Simultaneously, by observing stellar-mass compact binary coalescences within this frequency range, it can also place precise constraints on cosmological parameters. 
The articles \citep{2009PhRvD..80j4009C, 2001PhRvL..87v1103S,2012PhRvD..85d4047N,2019EPJC...79..900Z}
have discussed in detail the capability of such detectors to probe cosmological parameters, including the Hubble constant and dark energy, using different detection methods. Below, we consider 
\citet{2009PhRvD..80j4009C} as an example for illustration.

For a \ac{BNS} merger event, if the detector's low-frequency cutoff is $f_{\rm low}$, the duration from when the event becomes detectable by the instrument until the binary coalescence is
\begin{equation}
t=0.86 \ {\rm day}\times(2 \ {\rm Hz}/f_{\rm low})^{8/3}.
\end{equation}
It is clear that for a given \ac{BNS} merger, BBO could, in principle, observe the signal for years or even decades. Considering the effect of the detector's motion on the instrument response function, GW sources can be precisely localized. Therefore, it is expected that their host galaxies can be identified through cross-matching with optical galaxy catalogs, thereby obtaining redshift information.
Consequently, for such detectors, the number of GW sources that can serve as standard sirens will increase dramatically, reaching the order of $10^6$. Furthermore, due to the low instrumental noise and long integration times, the redshift distribution of detectable sources will be greatly extended, with the capability to detect sources up to $z\sim5$ or even higher. This will significantly enhance the ability to probe various cosmological parameters. 
\citet{2009PhRvD..80j4009C}
suggest that BBO is expected to constrain the Hubble constant to an accuracy of $0.1\%$. The constraints on dark energy parameters could reach $\Delta w_0\sim 0.01$ and $\Delta w_a\sim 0.1$.

\section{Examples of Typical Standard Sirens}
\label{sec5}

In this section, we focus on discussing three important types of GW sources as standard sirens, the main challenges associated with these methods, and their constraining abilities on cosmological parameters.

\subsection{\ac{BNS} Merger GW Sources and Their Electromagnetic Counterpart Observations} 
\label{sec5.1}

\begin{figure}[!htb]
\begin{center}
\includegraphics[scale=0.3]{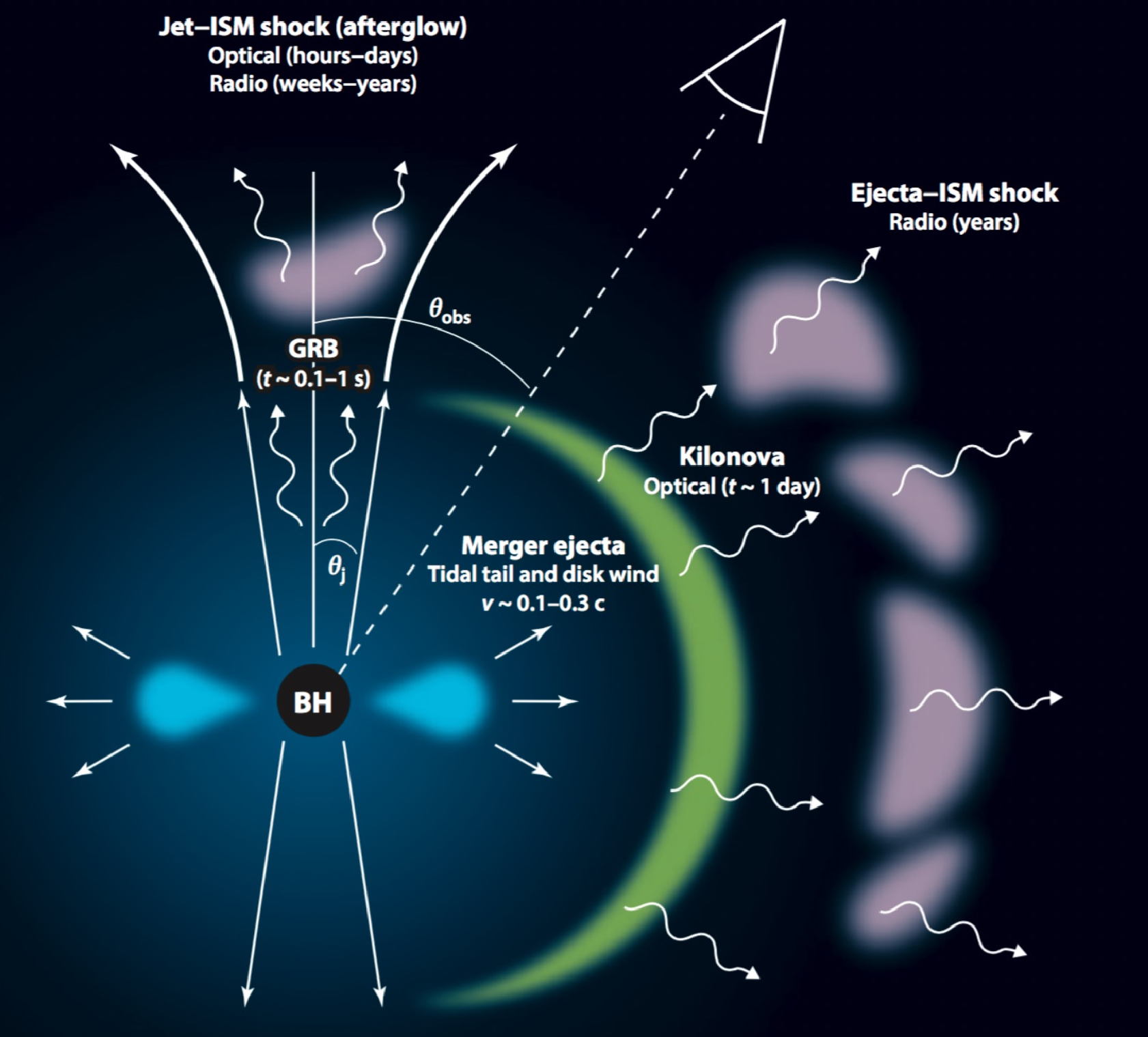}
\caption{This figure presents a
schematic diagram for various
electromagnetic counterparts of a \ac{BNS} merger \citep{2012ApJ...746...48M, 2020LRR....23....1M}.}
\label{b1}
\end{center}
\end{figure}

As mentioned above, the most suitable GW sources for use as standard sirens are \ac{BNS} mergers or NSBH merger events. Due to the presence of neutron stars, such events are typically accompanied by multi-wavelength electromagnetic radiation. 
Currently, through observations of the electromagnetic counterparts of GW170817, it has been firmly established that these GW bursts are accompanied by nearly isotropic kilonova emissions, as well as short gamma-ray bursts with their optical and radio afterglows \citep{2017ApJ...848L..12A, 2021ARA&A..59..155M}. The relative positions of these various electromagnetic radiation components with respect to the remnant black hole following a \ac{BNS} merger are illustrated in Figure \ref{b1}. 
Furthermore, some theories and observations suggest that such sources may also be accompanied by isotropic X-ray plateaus. Therefore, for these sources, simultaneous measurement of their luminosity distance and redshift can be achieved through joint observations of GWs and multi-wavelength electromagnetic radiation, enabling their use as standard sirens. Since the models for electromagnetic counterparts of NSBH merger GW events are not yet well established \citep{2026ApJ..1000L..20L}, and their radiation intensity depends on many factors (such as the black hole's mass and spin, the neutron star's equation of state, etc.), this section will focus solely on GW sources from \ac{BNS} mergers.

The observations on their electromagnetic counterparts can be broadly divided into two categories. For relatively nearby sources with redshifts $z<0.1$, their kilonova emission is expected to be directly observable by upcoming optical telescopes, including ZTF \citep{2019PASP..131g8001G, 2019PASP..131f8003B}, 
WFST \citep{2023SCPMA..6609512W}, LSST \citep{2019ApJ...873..111I}. Through their optical counterparts, the host galaxy can be directly identified by crossmatching with galaxy catalogs, allowing for a direct redshift measurement. In this case, since kilonova emission is nearly isotropic, the electromagnetic counterparts for the vast majority of events are likely to be directly detected. Given these characteristics, these nearby sources can be used to precisely measure the Hubble constant and hold promise for resolving the current tension in Hubble constant measurements.

However, for more distant sources, the kilonova emission is too faint to be detected directly. Therefore, the primary electromagnetic counterparts anticipated are short gamma-ray bursts and their afterglow radiations. Since both gamma-ray bursts and their afterglows are relatively brighter, they can potentially be detected at very large distances, making them well-suited for measuring the equation of state of dark energy. 
Their drawback, however, is also significant. Current observations suggest that short gamma-ray bursts and their afterglows are beamed emissions, with a typical jet opening angle of about $5^\circ$ \citep{2018Natur.561..355M}. Consequently, only sources whose jets are oriented almost directly towards the observer are likely to have detectable electromagnetic counterparts, representing only a very small fraction of all such events. 
Here, we focus on the latter scenario. We analyze the detection capabilities for these sources in the era of third-generation GW detectors, either by a single detector or by a network of detectors. We also examine the probability and spatial distribution of detecting the accompanying short gamma-ray bursts and their optical afterglows. Finally, we discuss the constraining abilities on dark energy parameters based on these sources. This section primarily references \cite{2021ApJ...916...54Y}.

\subsubsection{\ac{BNS} Populations and Their Gravitational Wave Detection}

The redshift evolution of the \ac{BNS} merger rate density can be calculated using the following formula:
\begin{equation}
	N(z)dz=\frac{R_{\mathrm{BNS ~mergers},0}\times f(z)}{1+z}\frac{dV(z)}{dz}dz,
\label{nz}
\end{equation}
where $R_{\mathrm{BNS mergers},0}$ is the local merger rate.
$f(z)$ is the redshift evolution function, and $dV(z)/dz$ is the differential comoving volume.
The function $f(z)$ depends on the initial distribution of \ac{BNS}s and the merger delay time. Here, we assume the initial distribution traces the star formation rate (SFR), which can be expressed as
\begin{equation}
	\mathrm{SFR}(z)\propto\left [(1+z)^{3.4\eta}+\left (\frac{1+z}{5000}\right )^{-0.3\eta}+\left (\frac{1+z}{9}\right )^{-3.5\eta}\right ]^{1/\eta},
\label{sfr}
\end{equation}
with units of $M_{\odot}\ \mathrm{Gpc}^{-3} \cdot \mathrm{yr}^{-1}$ and parameter $\eta=-10$. For the delay time distribution, we adopt a log-normal model. In this model, the distribution of delay times $\tau$ is
\begin{equation}
	P(\tau)d\tau=\frac{1}{\sqrt{2\pi}\tau\sigma}\exp\left [-\frac{\left (\ln\tau-\ln t_{d}\right )^{2}}{2\sigma^{2}}\right ]d\tau,
\label{tau}
\end{equation}
where $t_{d}=2.9\ \mathrm{Gyr}$ and $\sigma=0.2$. Therefore, the overall merger rate density at a given redshift can be computed via the following integral:
\begin{equation}
	f(z(t_{0}))=\int \mathrm{SFR}(z(t_{0}-\tau))P(\tau)d\tau.
\label{fz}
\end{equation}

In this section, we consider both second-generation and third-generation GW detectors. The former includes the network LHVIK, comprising LIGO (Livingston), LIGO (Hanford), Virgo, KAGRA, and LIGO-India. The latter includes ET, CE, and a CE-type detector located in Australia. Additionally, we consider a 2.5-generation detector network, LHV A+ and LHVIK A+. For each GW source, we define it as detectable if its signal-to-noise ratio (SNR) is greater than 12; otherwise, it is considered undetectable.

\subsubsection{Electromagnetic Counterparts}

As mentioned above, in this section we consider two types of electromagnetic counterparts: short gamma-ray bursts and their optical afterglows.

$\bullet$ ~~ \emph{Short Gamma-Ray Bursts}

Multi-wavelength observations of the GW event GW170817/GRB 170817A support a Gaussian-structured jet model. The jet profile can be expressed as: for $\iota \leq \iota_{w}$,
\begin{equation}
	E(\iota)=E_{0}\exp\left (-\frac{\iota^{2}}{2\iota_{c}^{2}}\right )
\label{jet}
\end{equation}
where $\iota_{c}= 0.057^{+0.025}_{-0.023}$, $\log_{10}E_{0}=52.73^{+1.3}_{-0.75}$, and $\iota_{w}=0.62^{+0.65}_{-0.37}$. In our calculations, we assume all \ac{BNS} merger events have such structured jets. To convert this energy profile into a luminosity profile, we assume the GRB duration is $T_{90}\sim 2$ s and the energy spectrum is uniformly distributed over this time. Therefore, the gamma-ray burst energy flux can be written as:
\begin{equation}
	F_{\gamma}=\frac{E_{0}\eta_{\gamma}}{4\pi D_{L}^{2} T_{90}}\exp\left (-\frac{\iota^{2}}{2\iota_{c}^{2}}\right )
\label{flux},
\end{equation}
where $\eta_{\gamma}$ is the radiative efficiency in the 1 – 10$^4$ keV band, which we set to $10\%$.

We consider several high-energy gamma-ray or X-ray telescopes: 
Fermi-GBM \citep{2009ApJ...702..791M}, with a sensitivity of $\sim2\times10^{-7}\ \mathrm{erg}\cdot \mathrm{s}^{-1}\cdot \mathrm{cm}^{-2}$ in the 50-300 keV band; 
GECAM  \citep{2022RDTM....6...12L}, with a sensitivity of $\sim1\times10^{-7}\ \mathrm{erg} \cdot \mathrm{s}^{-1} \cdot \mathrm{cm}^{-2}$ in the 50-300 keV band; 
Swift-BAT  \citep{2005SSRv..120..143B} and SVOM-ECLAIRS \citep{2009AIPC.1133...25G}, both with a sensitivity of $\sim1.2\times10^{-8}\ \mathrm{erg}\cdot \mathrm{s}^{-1} \cdot \mathrm{cm}^{-2}$ in the 15-150 keV band; 
and finally Einstein Probe (EP) \citep{2025SCPMA..6839501Y}, with a sensitivity of $\sim3\times10^{-9}\ \mathrm{erg} \cdot \mathrm{s}^{-1} \cdot \mathrm{cm}^{-2}$ in the 0.5-4 keV band. 
Another factor is the telescope's field of view (FoV). Fermi-GBM has an FoV covering $\sim3/4$ of the sky, GECAM covers the full sky, Swift-BAT covers $\sim1/9$ of the sky, SVOM-ECLAIRS covers $\sim1/5$ of the sky, and EP covers $\sim 1/11$ of the sky. 

$\bullet$ ~~ \emph{Optical Afterglows}

We consider the standard afterglow model. The spectrum depends on the viewing angle $\iota$, the jet half-opening angle $\theta_j$, the total isotropic-equivalent kinetic energy $E_j=(1-\cos\theta_j)E_0$, the ambient medium number density $n_0$, the magnetic field energy fraction $\epsilon_B$, the accelerated electron energy fraction $\epsilon_e$, the shock-accelerated electron spectral index $p$, the luminosity distance $d_L$, and the time since merger $t_j$.

Here, we adopt the shorthand notation $Q\equiv10^{x}Q_{,x}$. For an adiabatic jet, the Lorentz factor is:
\begin{equation}
	\gamma(t)=8.9(1+z)^{3/8}E_{j,51}^{1/8}n_{0}^{-1/8}\theta_{j, -1}^{-1/4}t_{,d}^{-3/8}.
\label{gamma_t}
\end{equation}
For an on-axis observer, a jet break occurs when $\gamma$ falls below $\theta_{j}^{-1}$. At the jet break time $t_j$, there is a break in the afterglow light curve. From Equation (\ref{gamma_t}), the corresponding jet break time $t_{j}$ is
\begin{equation}
	t_{j}=0.82(1+z)E_{j, 51}^{1/3}n_{0}^{-1/3}\theta_{j, -1}^{2}\ \mathrm{days}. 
\end{equation}
At the jet break time, there are two different spectral regimes: the slow-cooling case ($\nu_{m}<\nu<\nu_{c}$), where the flux density scales as $F_{\nu, j}\propto \nu^{-(p-1)/2}$ with $p=2.2$; and the fast-cooling case ($\nu_{c}<\nu$), where $F_{\nu, j}\propto\nu^{-p/2}$.

For an on-axis observer, the light curve is divided into two power-law segments by the jet break time. We denote the temporal decay indices of the flux density $F_{\nu,0}(t)$ as $\alpha_{1}$ and $\alpha_{2}$. For $t<t_j$, in the fast-cooling regime $\alpha_1=-(2-3p)/4$, while in the slow-cooling regime $\alpha_1=-3(1-p)/4$. However, for $t>t_{j}$, in the isotropic fireball model, the on-axis observer only sees a fraction of the flux. Using the relation $\gamma(t)\propto t^{-3/8}$, we have $\alpha_{2}=\alpha_{1}+3/4$.

For the detection of optical afterglows, we consider two optical telescopes. The first is LSST, with a $5\sigma$ point-source limiting magnitude of 24.7 in the $r$-band \citep{2019ApJ...873..111I}. The second is WFST, with a limiting magnitude of 22.8 in the $r$-band for a 30-second exposure and 24.1 for a 300-second exposure \citep{2023SCPMA..6609512W}.

\subsubsection{Multi-messenger Detections}

\begin{figure}[!htb]
\begin{center}
\includegraphics[scale=0.45]{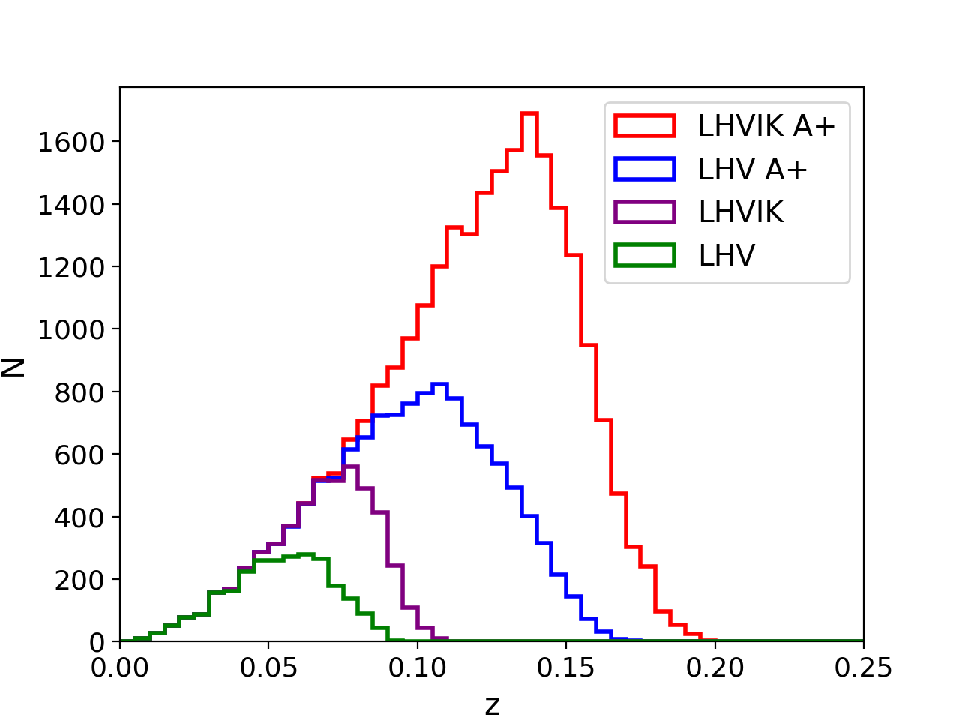}
\includegraphics[scale=0.45]{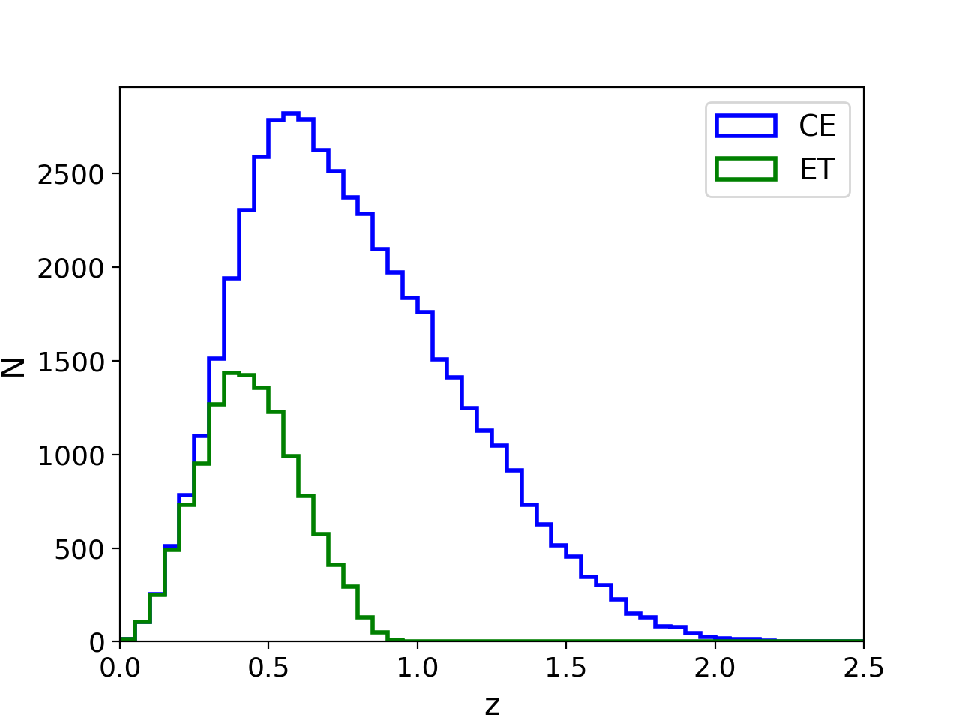}
\caption{Redshift distribution of GW events jointly detected by GW detectors and the GECAM detector.}
\label{a8}
\end{center}
\end{figure}

Through numerical simulations, we investigate the event rates for the joint detection of GWs, gamma-ray bursts, and optical afterglows by their respective detectors. Combining these with the current \ac{BNS} merger rate, we obtain the annual joint detection rates for various GW detector networks and high-energy detectors, which are listed in Table \ref{table1}. Based on the corresponding simulation results, we draw the following conclusions:
\begin{itemize}
  \item{(1)} For all \ac{BNS} merger events detectable by both GW and high-energy detectors, their optical afterglows are sufficiently bright to be detected by either LSST or WFST; 
  \item{(2)} In the era of second-generation detectors, the joint detection rate is very low, and the redshifts of these events are mostly distributed below $z \lesssim 0.2$ (see Figure \ref{a8}), making them unsuitable for dark energy studies;
  \item{(3)} Wide-field instruments like GECAM and Fermi-GBM are more suitable for joint observations compared to smaller field-of-view telescopes such as EP;
  \item{(4)} For all events that can be jointly detected, the viewing angle satisfies $\iota \lesssim 20^{\circ}$, meaning they are observed nearly on-axis;
  \item{(5)} In the era of third-generation detectors, the joint detection rate can increase substantially, reaching an annual rate on the order of tens to thousands. The redshift distribution of these events is primarily concentrated in the range $z \sim 0.5-2$ (see Figure \ref{a8}), making them effective for probing the dark energy equation of state and its evolution;
  \item{(6)} For joint detection, CE is more suitable than ET due to its lower noise curve. 
\end{itemize}

\begin{table}[t]
\caption{Annual event rate for multi-messenger observations. }
\centering
\begin{tabular}{|c|c|c|c|c|c|}
\hline
&Swift-BAT&SVOM-ECLAIRS&GECAM&Fermi-GBM&EP\\
\hline
LHV&0.042-0.425 &0.072-0.731&0.278-2.820&0.198-2.001&0.029-0.297\\
\hline
LHVIK&0.084-0.856&0.146-1.474&0.553-5.598&0.394-3.985&0.058-0.593\\
\hline
LHV A+&0.217-2.200&0.374-3.789&1.370-13.870&0.962-9.741&0.148-1.504\\
\hline
LHVIK A+&0.445-4.505&0.766-7.757&2.743-27.769&1.907-19.305&0.301-3.046\\
\hline
ET&17.0-172.0&29.2-296.1&80.6-815.6&49.9-504.9&10.7-108.5\\
\hline
CE&98.1-993.0&168.9-1710.0&342.1-3463.5&188.4-1907.4&58.1-587.9\\
\hline
\end{tabular}
\label{table1}
\end{table}

Using these simulated GW events, we can calculate their constraining ability on the dark energy parameters $(w_0, w_a)$. Here, we consider only third-generation GW detector scenarios: a single ET, a single CE, an ET+CE network, and an ET+CE+CE-type network. 
For each simulated GW event, we assume its redshift can be precisely determined from the electromagnetic counterpart, while the error on the luminosity distance is derived from the GW data. The corresponding errors on the dark energy parameters are given in Table \ref{table2}. We find that CE can provide tighter constraints than ET. 
In the most optimistic scenario, joint GW observations could potentially constrain the error on $w_0$ to the level of $0.01$ and the error on $w_a$ to the level of $0.1$, which meets or even exceeds the projected detection limits of next-generation traditional methods like supernovae. Therefore, the GW standard siren method can fully serve as one of the primary tools for future dark energy exploration. 
Finally, we must note that this method also has limitations. In GW observations, there exists a strong degeneracy between the luminosity distance and the source's viewing angle. In the calculations above, we had to assume that future electromagnetic observations can accurately determine the viewing angle. However, this method of determining the viewing angle relies on models of the electromagnetic emission from GW events. 
For sources with viewing angles greater than $\sim 5^{\circ}$, this method carries a degree of model dependence, introducing some uncertainty into future observations. How to better break this degeneracy is currently a topic of active research.

\begin{table*}
\caption{Potential constraints on dark energy parameters from one year of joint observations. }
\begin{center}
\begin{tabular}{|c|c|c|c|c|c|c|}
\hline
\multicolumn{2}{|c|}{}&Swift-BAT&SVOM-ECLAIRS&GECAM&Fermi-GBM&EP\\
\hline
&ET&0.129-0.412&0.099-0.314&0.057-0.181&0.070-0.222&0.160-0.510\\
\cline{2-7}
$\Delta w_0$&CE&0.034-0.107&0.026-0.082&0.016-0.051&0.020-0.065&0.043-0.136\\
\cline{2-7}
&CEET&0.032-0.104&0.025-0.079&0.015-0.049&0.020-0.062&0.041-0.131\\
\cline{2-7}
&CE2ET&0.028-0.090&0.022-0.069&0.013-0.042&0.017-0.054&0.036-0.114\\
\hline
&ET&1.173-3.734&0.894-2.846&0.531-1.690&0.664-2.114&1.462-4.652\\
\cline{2-7}
$\Delta w_a$&CE&0.237-0.754&0.181-0.575&0.119-0.380&0.157-0.499&0.303-0.966\\
\cline{2-7}
&CEET&0.228-0.727&0.174-0.554&0.115-0.366&0.151-0.481&0.293-0.931\\
\cline{2-7}
&CE2ET&0.198-0.630&0.151-0.480&0.100-0.318&0.132-0.419&0.254-0.808\\
\hline
\end{tabular}
\end{center}
\label{table2}
\end{table*}

\subsection{\ac{SBBH} Mergers as Dark Sirens} 
\label{sec5.2}

Unlike \ac{BNS} mergers, most \acp{SBBH} mergers are unlikely to have observable electromagnetic counterparts. To measure their redshifts, a common method is to utilize the spatial localization capability of GW detector networks to search for the host galaxies of the \acp{SBBH}. The standard sirens obtained through this method, which do not rely on electromagnetic counterparts, are often referred to as ``dark sirens''. However, due to the relatively poor angular resolution of GW detectors, the localization volume for a given event typically contains thousands of potential host galaxies, posing a significant challenge for host identification. 
In 
\cite{2020MNRAS.498.1786Y}
proposed a method to directly search for the host galaxy group of a GW source by cross-matching the GW localization volume with galaxy group catalogs, rather than searching for individual host galaxies. This approach offers several advantages: (1) Compared to individual galaxies, galaxy groups represent larger-scale physical structures, making the host group of a GW event easier to identify; (2) For a given galaxy survey, the completeness of a galaxy group catalog is typically much higher than that of a galaxy catalog; (3) Measuring the source redshift via its host group can partially avoid the impact of galaxy peculiar velocities on redshift measurements, as the group-finding method helps average over the virial motions of galaxies within their dark matter halos, reducing adverse effects on cosmological parameter inference. 
Based on these advantages, this section investigates the capability of various detector networks to resolve the host galaxy groups of \ac{SBBH} merger events and discusses the application of such dark sirens in constraining the Hubble constant. 

\subsubsection{Searching for Host Galaxy Groups}

In this section, we employ an SDSS galaxy group catalog. During its construction, all SDSS DR7 galaxies within the redshift range $0.01 \leq z \leq 0.20$ and with an edge-completeness factor $f_\mathrm{edge} > 0.7$ were included in the group-finding algorithm. Galaxy colors and magnitudes in the catalog are derived from standard SDSS Petrosian techniques and corrected for Galactic extinction. For each galaxy group in the catalog, its halo mass $M_h$ is assigned via abundance matching based on the group's characteristic stellar mass $M_{*,\ \mathrm{halo}}$, defined as the sum of the stellar masses of all member galaxies brighter than $M_r = -19.5$ in the $r$-band.
We select massive galaxy groups with $\log_{10}[M_{\mathrm{halo}}/(M_\odot h^{-1})] > 12$. Similar to the case of host galaxy searches, to ensure catalog completeness, we ignore the possibility of low-mass halos hosting the GW source. Based on the halo mass, we can also estimate the spatial distribution of member galaxies, which is influenced by the virial radius $R_h$ of the host dark matter halo. $R_h$ is given by
\begin{equation}
R_h = \left( \frac{3M_\mathrm{halo}}{4\pi \delta \bar{\rho}} \right)^{1/3},
\label{5.3}
\end{equation}
where $\bar{\rho}$ is the mean density of the universe, and $\delta = 180$, indicating that we define the boundary of the dark matter halo as the region where the density exceeds $180\bar{\rho}$.
Furthermore, we must consider two factors affecting the catalog completeness. First, the survey footprint has a specific geometry. Therefore, the projected area of some galaxy groups may intersect one or more survey edges, potentially causing member galaxies to fall outside the observed region and leading to incompleteness. This would affect the accuracy of identifying the host group of a GW source. 
To mitigate this effect, we select groups with an edge-completeness factor $f_\mathrm{edge} > 0.7$, which we treat as being fully contained within the survey footprint. The second factor arises from instrumental limitations. At higher redshifts, due to the flux limit of the SDSS survey, some galaxies and groups may be missed. An analysis of the spatial distribution of group counts reveals that the catalog is complete for $z < 0.12$. Therefore, as a conservative approach, in the following discussion we only consider low-redshift groups within $z < 0.1$. In this region, the number density of galaxy groups is $2\times10^{-3}\ \mathrm{Mpc}^{-3}$.

To investigate the capability of identifying host galaxy groups, we assume that for each galaxy group in the catalog, a \ac{SBBH} merger event occurs at its center $(\alpha,\delta,\log(d_L))$, and perform hypothetical observations of these events with different GW detector networks. For these simulated GW events, we calculate the errors and covariance matrix for all parameters using the Fisher matrix formalism, which is then marginalized down to the covariance matrix for the three parameters $(\alpha,\delta,\log(d_L))$. 
Subsequently, we use this covariance matrix to compute the three-dimensional error ellipsoid for each GW event (we consider the 99\% confidence localization volume) and count the number of galaxy groups from the catalog contained within this volume. This count represents the number of candidate host groups for that event. If only one galaxy group lies within the error ellipsoid, we consider the host group to be uniquely identified for that event. However, if multiple groups are present, we must statistically analyze these groups to estimate the redshift distribution of the GW source.

\begin{figure}[!htb]
\begin{center}
\includegraphics[scale=0.6]{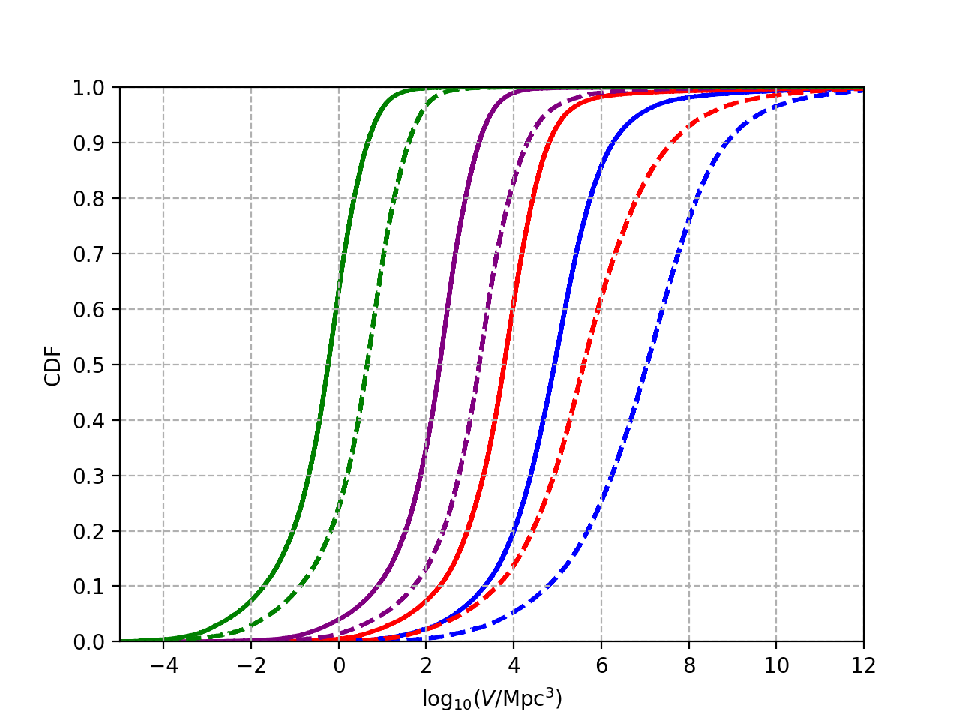}
\caption{Cumulative probability distribution of the localization error volume for $30$–$30\,M_\odot$ \ac{SBBH}s (solid lines) and $10$–$10\,M_\odot$ \ac{SBBH}s (dashed lines) with different detector networks. Blue, red, purple, and green lines correspond to LHV, LHVIK, LHVIKCA, and CE2ET, respectively.}
\label{b2}
\end{center}
\end{figure}

The localization capability of a given detector network depends on the parameters of the merging system, particularly the black hole masses and the orbital inclination angle $\iota$. For the simulated \ac{SBBH} mergers, based on current GW observations, we consider two specific mass pairs: $30-30\ M_\odot$ and $10-10\ M_\odot$. For the orbital inclination, we assume a distribution proportional to $\sin\iota$. 
For the detector networks, we consider three 2G/2.5G configurations: LHV, LHVIK, and LHVIKCA. Note that ``C'' and ``A'' here denote proposed 8-km detectors located in China and Australia, respectively. For 3G detectors, since a network consisting of only two detectors provides poor sky localization (e.g., CEET's localization capability is far inferior to LHVIKCA), we only discuss the CE2ET network here. 
In Figure \ref{b2}, we show the cumulative distribution function (CDF) of the localization volume. We find that for the 2G network LHVIK, the typical localization volume is $(10^3-10^5)\ \mathrm{Mpc}^3$, while for the 3G network CE2ET, this volume is dramatically reduced to $(0.1-10)\ \mathrm{Mpc}^3$.

\begin{figure}[!htb]
\begin{center}
\includegraphics[scale=0.6]{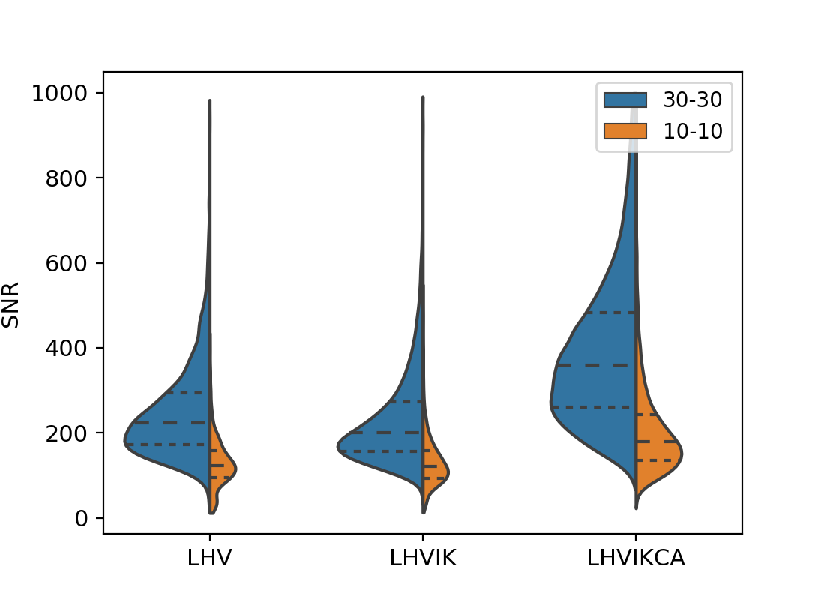}
\caption{Violin plot of the SNR for \ac{SBBH} merger events with $N_\mathrm{in}=1$. 
From left to right, the results are for LHV, LHVIK, and LHVIKCA, respectively. In each violin plot, the left part corresponds to $30$–$30\,M_\odot$ \ac{SBBH}s, and the right part corresponds to $10$–$10\,M_\odot$ \ac{SBBH}s. The area of each side within a given violin plot is proportional to the number of SBBHs with $N_\mathrm{in}=1$. 
Note that due to the large difference in the number of events between LHV and LHVIKCA, the ratio between area and event count varies significantly across different violin plots for display purposes; therefore, direct comparisons between different violin plots are not valid.}
\label{b3}
\end{center}
\end{figure}

In Figure \ref{b3}, we present violin plots showing the signal-to-noise ratio (SNR) distribution for \ac{SBBH} merger samples with $N_\mathrm{in}=1$ for the LHV, LHVIK, and LHVIKCA networks. The left and right sides of each violin plot correspond to the results for $30-30\ M_\odot$ and $10-10\ M_\odot$ binaries, respectively. It can be seen that the SNR for $30-30\ M_\odot$ binaries is generally higher than that for $10-10\ M_\odot$ \acp{SBBH}. In Table \ref{table_5.3}, we list the fractions of \ac{SBBH} merger events for various scenarios: $N_\mathrm{in}=1$, $N_\mathrm{in}\leq2$, $N_\mathrm{in}\leq5$, and $N_\mathrm{in}\leq10$.

\begin{table}
\caption{Distribution proportion of $N_{\mathrm{in}}$ in different scenarios and the corresponding event rates.}
\begin{center}
\begin{tabular}{cccccc}
\hline\hline
&$N_\mathrm{in}$&1&$\leq2$&$\leq5$&$\leq10$\\
\hline
LHV&$30-30\ M_\odot$&5.29\%&8.23\%&13.79\%&20.05\%\\
&&(0.2-2.2\ yr$^{-1}$)&(0.3-3.4\ yr$^{-1}$)&(0.5-5.6\ yr$^{-1}$)&(0.8-8.2\ yr$^{-1}$)\\\\
&$10-10 M_{\odot}$&1.60\%&2.41\%&3.83\%&5.24\%\\
&&(0.06-0.6\ yr$^{-1}$)&(0.09-1.0\ yr$^{-1}$)&(0.2-1.6\ yr$^{-1}$)&(0.2-2.1\ yr$^{-1}$)\\
\\
LHVIK&$30-30\ M_{\odot}$&17.00\%&26.33\%&44.41\%&62.05\%\\
&&(0.7-6.9\ yr$^{-1}$)&(1.0-10.7\ yr$^{-1}$)&(1.7-18.1\ yr$^{-1}$)&(2.4-25.3\ yr$^{-1}$)\\\\
&$10-10\ M_{\odot}$&5.04\%&7.24\%&11.59\%&17.62\%\\
&&(0.2-2.0\ yr$^{-1}$)&(0.3-3.0\ yr$^{-1}$)&(0.4-4.7\ yr$^{-1}$)&(0.7-7.2\ yr$^{-1}$)\\
\\
LHVIKCA&$30-30\ M_{\odot}$&59.17\%&77.89\%&93.55\%&98.16\%\\
&&(2.3-24.1\ yr$^{-1}$)&(3.0-31.8\ yr$^{-1}$)&(3.7-38.2\ yr$^{-1}$)&(3.8-40.0\ yr$^{-1}$)\\\\
&$10-10\ M_{\odot}$&30.75\%&46.29\%&69.86\%&83.88\%\\
&&(1.2-12.5\ yr$^{-1}$)&(1.8-18.9\ yr$^{-1}$)&(2.7-28.5\ yr$^{-1}$)&(3.3-34.2\ yr$^{-1}$)\\
\\
CE2ET&$30-30\ M_{\odot}$&99.01\%&99.92\%&99.99\%&100\%\\
&&(3.9-40.4\ yr$^{-1}$)&(3.9-40.8\ yr$^{-1}$)&(3.9-40.8\ yr$^{-1}$)&(3.9-40.8\ yr$^{-1}$)\\\\
&$10-10\ M_{\odot}$&94.36\%&99.06\%&99.93\%&99.99\%\\
&&(3.7-38.5\ yr$^{-1}$)&(3.9-40.4\ yr$^{-1}$)&(3.9-40.8\ yr$^{-1}$)&(3.9-40.8\ yr$^{-1}$)\\
\hline
\\
\end{tabular}
\end{center}
\label{table_5.3}
\end{table}

\subsubsection{Constraining the Hubble Constant}

To calculate the constraints on the Hubble constant, we randomly selected 100 galaxy groups from the catalog and placed a \ac{SBBH} merger event at the center of each, at the coordinates $(\alpha,\delta,d_L(z) )$. For each simulated event, we derived the error ellipsoid for the spatial parameters $(\alpha,\delta,\log(d_L))$ using Fisher matrix analysis. Subsequently, we traversed all galaxy groups within the 99\% confidence error ellipsoid, assigning each a weight factor of 1. Using Bayesian inference, we then computed the posterior probability distribution for the Hubble constant $H_0$.
In our simulations, we found that for the LHV network, due to its very large $\Delta d_L/d_L$ values, almost all simulated events were discarded because their error ellipsoids extended beyond the survey boundaries. Therefore, we could not obtain a posterior distribution for $H_0$ with the LHV configuration. For this reason, we only discuss the results for the LHVIK, LHVIKCA, and CE2ET detector networks.

For the 2G network LHVIK, among the 100 simulated GW events, 17 GW events from the $30-30\ M_\odot$ SBBH sample and only 3 GW events from the $10-10\ M_\odot$ SBBH sample met the selection criterion. In Table \ref{table_5.5}, we list the constraining ability on $H_0$ from such observations.
For the LHVIKCA network, 44 $30-30\ M_\odot$ SBBH mergers satisfied the selection criteria and could be used to constrain $H_0$. Compared to the LHVIK results, the uncertainty on $H_0$ is reduced by approximately a factor of five. Due to the contribution of the two proposed 8-km detectors, the localization capability of the LHVIKCA network for individual events is significantly improved. This enhancement impacts the $H_0$ constraint in two ways. First, the redshift uncertainty for GW events is greatly reduced. Second, the number of events meeting the selection criteria increases. Therefore, we conclude that if plans for 8-km detectors materialize in the near future, the utility of SBBH dark standard sirens for cosmology in the 2G detector era would be substantially improved.
For the 3G detector network CE2ET, as expected, we find the precision on $H_0$ is much better to that from 2G networks. 
For the $30-30\ M_\odot$ SBBH mergers, 66 events could be used for parameter constraints, under different prior selections for galaxy groups, the expected best results in both cases were $\Delta H_0/H_0 \sim 0.008\%$. 
For the $10-10\ M_\odot$ SBBH mergers, 65 effective events yielded an expected best result of $\Delta H_0/H_0 \sim 0.018 \%$ .

\begin{table}
\caption{Constraints on the Hubble constant from different detector arrays. 
The third column shows the fraction of effective events in a sample of 100 random \ac{SBBH} mergers. The fourth column presents the constraint on $\Delta H_0/H_0$ as a function of the number of effective events $N$. The last column gives the $\Delta H_0/H_0$ result from five years of GW observations.   }
\centering
\begin{tabular}{ccccc }
	\hline\hline
	Network & BH mass&f$_\mathrm{effective}$&N effective events&Five-year observation\\
	\hline
	LHVIK & $30-30\ M_{\odot}$&17\%&~8.1\%/$\sqrt{N}$&1.4\%-4.4\%\\
	& $10-10\ M_{\odot}$&3\%&NA&NA\\\\
	LHVIKCA & $30-30\ M_{\odot}$&44\%&~2.5\%/$\sqrt{N}$&0.26\%-0.85\%\\
	& $10-10\ M_{\odot}$&21\%&~5.6\%/$\sqrt{N}$&0.86\%-2.74\%\\\\
	CE2ET & $30-30\ M_{\odot}$&66\%&~0.09\%/$\sqrt{N}$&~0.008\%-0.026\%\\
	& $10-10\ M_{\odot}$&65\%&~0.21\%/$\sqrt{N}$&~0.018\%-0.057\%\\
	\hline\\
\end{tabular}
\label{table_5.5}
\end{table}

\subsection{Dark lensed sirens}

Lensed GW events are expected to be powerful cosmological probes via their time-delay distance measurements as described earlier in Section~\ref{lensGW}, provided that their redshifts can be obtained from electromagnetic observations. However, many lensed GW events, especially mergers of SBBHs, may lack electromagnetic counterparts. \citet{2025ApJ...993L..57C} pointed out that the lensed host galaxies of these ``dark'' lensed GW events could be identified if the localization via GW signals is sufficiently accurate, albeit challenging. They showed that third-generation ground-based GW detectors will detect a population of lensed events with three or more detectable images (including the central image), each arriving at different times and at different locations of Earth in space. This effectively forms a network that reduces the typical localization area to $\sim0.01$\,deg$^2$. For at least $50\%$ of these events, the localization improves by more than a factor of $30$ compared with unlensed cases. Such precise localization enables robust host-galaxy identification by matching the lensed GW event with its lensed host galaxy discovered by large scale sky surveys, such as LSST, CSST, and Roman, in the localized area. The reason is that the lensed host galaxies for a large fraction of lensed GW events can be detected by these surveys and the average number of the lensed galaxies discovered by these surveys in a sky area of $0.1$\,deg$^2$ is roughly no more than $1$ \citep{2022ApJ...940...17C, 2025ApJ...993L..57C}. Once the lensed host galaxy of a dark lensed GW event is identified, then multiple-image detections, through lens modeling, further yield sub-arcsecond position. As dark lensed sirens, these events become powerful probes of cosmological parameters by the combination of both their time-delay distance and luminosity distance measurements. Using simulated lensed compact-binary mergers, they showed that about $2$\,yr or slightly longer observations with third-generation GW detectors can measure the Hubble constant to $\lesssim 1\%$ precision via ``dark lensed sirens'' (even when relying solely on lensed SBBH events) while simultaneously constraining other cosmological parameters. 

\section{Conclusions}
\label{sec7}

Various GW events discovered by the LIGO, Virgo, and KAGRA detectors, particularly the \ac{BNS} merger GW170817, have ushered in a new era of GW astronomy. For ground-based and space-based laser interferometer GW detectors, as well as \ac{PTA}s, the most significant sources are the inspirals and mergers of compact binaries.
These GW events occur across cosmic time,
thereby providing a novel pathway to study cosmology through observations of these sources and their electromagnetic counterparts. 
In this paper, we have reviewed the fundamental principle of using GW sources from compact binary coalescences as ``standard sirens'' to investigate the cosmic expansion history, which requires independent measurements of the source distance and redshift. 
We have detailed two methods for measuring distance and seven methods for determining source redshift mentioned in the current literature, comparing their advantages and disadvantages. 
For various ground-based and space-based GW detectors, we have discussed their potential to constrain cosmological parameters and models, particularly the Hubble constant and the dark energy equation of state. 
Research indicates that second-generation ground-based detectors, represented by Advanced LIGO, can be used to constrain the Hubble constant, potentially achieving a precision comparable to current optical methods. Meanwhile, third-generation GW detectors and space-based detectors are expected to precisely constrain dark energy. These results demonstrate the immense potential of GW sources as powerful probes for cosmology. In particular, they may act as one of the primary probes to clarify the issues of Hubble tension and/or dark energy evolution in the near future.

\normalem
\begin{acknowledgements}
This work is supported the National Natural Science Foundation of China (grant Nos. 12325301 12273035, 12273050, and 12533009), Strategic Priority Research Program of the Chinese Academy of Science (grant No. XDB0550300), the National Key R\&D Program of China (grant Nos. 2021YFC2203102 and 2022YFC2204602), and 
the National Astronomical Observatory of China (grant no. E4TG660101).
\end{acknowledgements}

\bibliographystyle{raa}
\bibliography{bibtex}

\end{document}